\documentclass[aps,pra,twocolumn,superscriptaddress,longbibliography]{revtex4-2}%
\usepackage{float}
\usepackage{graphics}
\usepackage[mathscr]{euscript}
\usepackage{amsfonts}
\usepackage{amssymb}
\usepackage{epsfig}
\usepackage{hyperref}
\usepackage{braket}
\usepackage{siunitx}
\usepackage{blkarray}
\usepackage{xcolor}
\usepackage{mathtools}
\usepackage{physics}
\usepackage[short,nocomma]{optidef}
\usepackage{multirow}
\usepackage{float}
\usepackage{chngcntr}
\usepackage{siunitx}
\usepackage[toc,page,header]{appendix}
\usepackage{minitoc}
\usepackage{soul}

\usepackage{mathtools}

\newcommand{\state}[1]{#1}

\newcommand{\hm}{H_\text{min}}

\newcommand{\totstd}{\sigma_t}
\newcommand{\totvar}{\sigma_t^2}
\newcommand{\noisevar}{\sigma_n^2}
\newcommand{\noisestd}{\sigma_n}
\newcommand{\vacvar}{\sigma_v^2}

\newcommand{\figref}[1]{Fig.~\ref{#1}}
\newcommand{\appref}[1]{App.~\ref{#1}}

\newcommand{\red}[1]{{\color{red}#1}}

\makeatletter
\def\blfootnote{\xdef\@thefnmark{}\@footnotetext}
\makeatother

\begin{document}

\blfootnote{MI, CRC, KNW, and CAB contributed equally to this work.}

\author{Makoto Ishihara}\affiliation{Department of Electronics and Electrical Engineering, Keio University, 3-14-1 Hiyoshi, Kohoku-ku, Yokohama 223-8522, Japan}
\author{Carles Roch i Carceller}\affiliation{Physics Department and NanoLund, Lund University, Box 118, 22100 Lund, Sweden}\affiliation{ICFO - Institut de Ciencies Fotoniques, The Barcelona Institute of Science and Technology, 08860 Castelldefels, Spain.}
\author{Kieran Neil Wilkinson}\altaffiliation[Now at ]{Quantinuum, Terrington House, 13-15 Hills Road, Cambridge, CB2 1NL, United Kingdom}\affiliation{Center for Macroscopic Quantum States (bigQ), Department of Physics, Technical University of Denmark, 2800 Kongens Lyngby, Denmark}
\author{Casper Ahl Breum}\altaffiliation[Now at ]{Novo Nordisk Fonden, Tuborg Havnevej 19, 2900 Hellerup, Denmark}\affiliation{Center for Macroscopic Quantum States (bigQ), Department of Physics, Technical University of Denmark, 2800 Kongens Lyngby, Denmark}
\author{Tobias Gehring}\affiliation{Center for Macroscopic Quantum States (bigQ), Department of Physics, Technical University of Denmark, 2800 Kongens Lyngby, Denmark}
\author{Jonatan Bohr Brask}\affiliation{Center for Macroscopic Quantum States (bigQ), Department of Physics, Technical University of Denmark, 2800 Kongens Lyngby, Denmark}

\title{Quantum randomness certification with untrusted measurements and few probe states}

\begin{abstract}
We present a scheme for semi-device-independent quantum randomness certification from an untrusted measurement device and a trusted source and demonstrate it experimentally. No assumptions about noise or imperfections in the measurement are required and the scheme is simple to implement with existing technology. The measurement device is probed with a few trusted states and the output entropy can be lower bounded conditioned on the observed outcome distribution. The protocol can be applied to measurements with any finite number of outcomes and in particular can be realised by homodyne measurements of the vacuum using a detector probed by coherent states, as we experimentally demonstrate by intensity modulation of a telecom-wavelength pilot laser followed by homodyne detection and discretisation by analog-to-digital conversion. We show that randomness can be certified in the presence of both Gaussian additive noise and non-Gaussian imperfections.
\end{abstract}
\maketitle

\section{Introduction}

Random numbers are central to a range of applications in science and technology, including numerical simulation, statistical sampling, gaming, and secure information processing \cite{Hayes2001}. In particular, for cryptographic applications, security relies on the inability of any adversary to predict the random numbers used to generate cryptographic keys. To establish security, it is thus crucial to provide rigorous bounds on the predictability of the randomness used.

Within classical physics, such randomness certification necessarily relies on assumptions on the knowledge and computational resources available to potential eavesdroppers. The inherent randomness in measurements on quantum systems, on the other hand, allows certification to be based directly on measurable properties of the devices used, under the reasonable assumption that adversaries are also constrained by quantum physics \cite{Acin2016,Herrero2017,Bera2017}. Provided a characterisation of a quantum state and measurement -- e.g.\ the output path of a single photon impinging on a beam splitter \cite{Stefanov2000} -- the predictability relative to any adversary can be bounded. Exploiting quantum nonlocality \cite{Bell1964,Brunner2014}, the need for accurate characterisation of the devices can even be eliminated, provided the measurement data violates a Bell inequality \cite{colbeckPhD2009,Pironio2010}. This has been demonstrated experimentally \cite{Pironio2010,Christensen2013,Bierhorst2018,Liu2018,Shalm2021,Liu2021} and provides a very strong level of security, known as device independence, since the devices may be largely untrusted.

However, device-independent schemes are also much more challenging to realise than simple device-dependent ones. It, therefore, makes sense to explore the trade-off between security and ease of implementation, identifying schemes that are fast and simple to realise at the price of introducing some limited amount of trust in the devices. Such protocols are broadly termed semi-device-independent, and many different setups with partially characterised state preparation or measurements have been considered, see e.g.~\cite{Li2011,Vallone2014,Lunghi2015,Chaturvedi2015,Cao2015,Nie2016,Cao2016,Xu2016,Nie2016,Marangon2017,Brask2017mhz,Avesani2018,Michel2019,Rusca2019,Drahi2020,Rusca2020,Mironowicz2021,Wang2023}. One simple approach to quantum randomness generation, which can reach very high rates, is to extract randomness from optical quadrature measurements of the vacuum state \cite{Gabriel2010,Symul2011,Zhang2016,Huang2019}. This also allows for fast, source-device-independent schemes \cite{Marangon2017,Avesani2018}, and security in settings with correlations across measurement rounds \cite{Gehring2021}.

\begin{figure}[t!]
	\includegraphics[width=\linewidth]{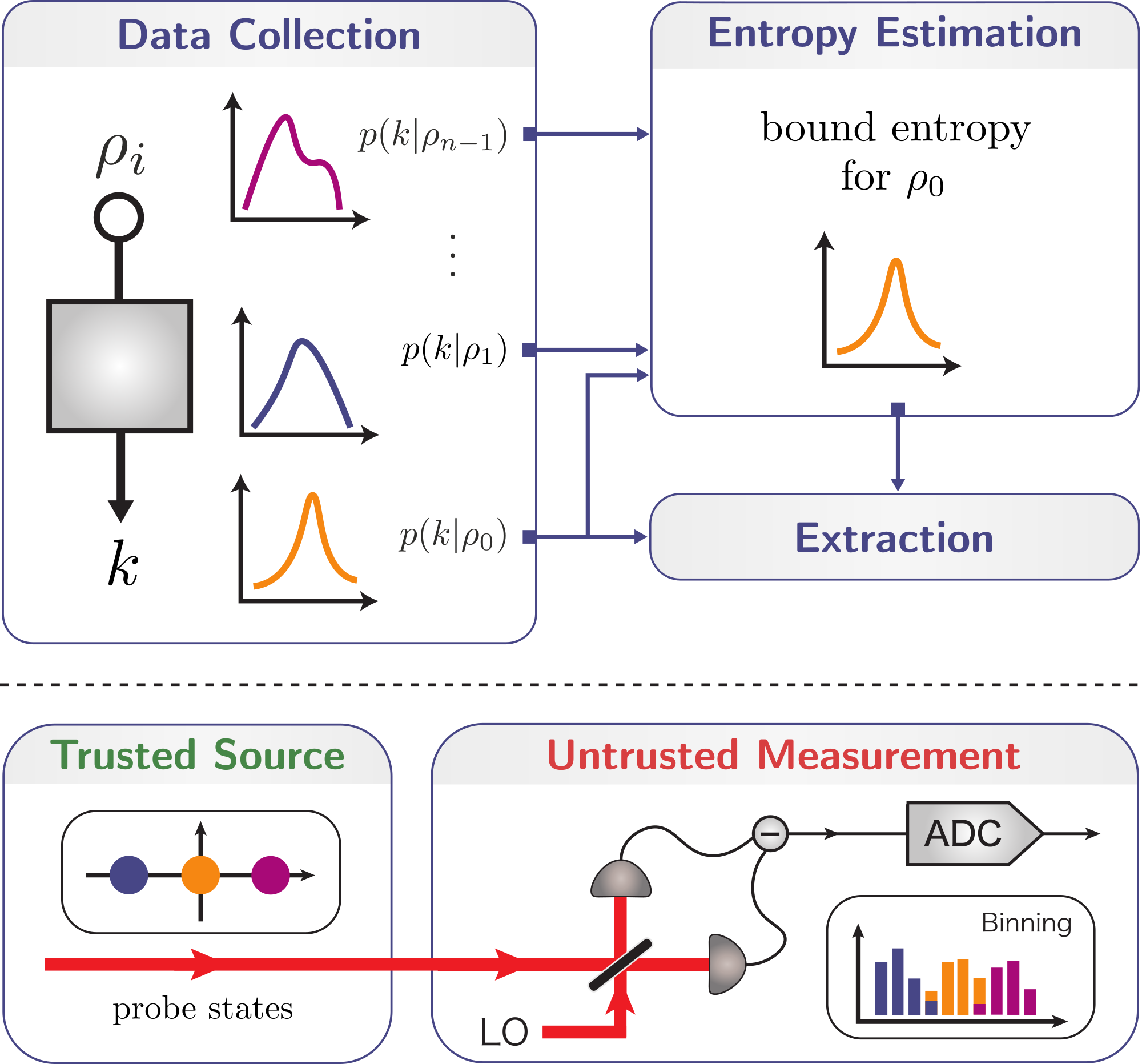}
	\caption{Top: General protocol. An untrusted detector is probed with a set of trusted states and the distributions of outcomes are recorded. The min-entropy of the data corresponding to one of the states is lower bounded given the observations, and finally randomness extraction is applied. Bottom: implementation based on homodyning the vacuum. The generation and additional probe states are the vacuum and a set of coherent states with varying amplitude and fixed phase. The measurement is homodyne detection. The signal is discretised into $d = 2^\Delta$ bins by an ADC. }
	\label{fig.concept}
\end{figure}

In this work, we first develop a method for randomness certification from a trusted source in prepare-and-measure setups, where the measurement is partially characterised using a small set of known input states. We then apply our framework to homodyning of the vacuum in an optical setup and demonstrate this scheme in an experiment.

The method is illustrated in the upper panel of \figref{fig.concept}. A measurement device -- initially an uncharacterised black box -- is probed using a (small) set of known quantum states $\{\rho_0,\ldots,\rho_{n-1}\}$, and the distributions $p(k|\rho_i)$ of outcomes are recorded. The entropy produced by measurements on a fixed state $\rho_0$ can be bounded via semidefinite programming using the observed data as constraints, without further assumptions about the behaviour or noise affecting the measurement device. This approach is general and applies to any prepare-and-measure scenario in which some input states may be trusted.

In the case of homodyne measurements of the vacuum, the fixed input is the vacuum state, and we take the remaining states to be coherent states with varying amplitude. The (untrusted) measurement is detection of a fixed quadrature, discretised by analogue-to-digital conversion to a finite number of outcomes. We show that significant randomness can be produced using just a few probe states, in the presence of both Gaussian and non-Gaussian noise. We note that the scheme operates far from the regime of detector tomography and already with the vacuum and a single coherent state, randomness can be extracted. Experimentally, we demonstrate our scheme in a simple optical setup with up to four probe states. In the following, we first describe the protocol and then our experimental implementation.

\section{Quantifying randomness}

The number of (almost) perfectly random bits relative to an adversary (Eve) that can be extracted from given raw data is quantified by its conditional min-entropy, according to the quantum left-over hash lemma \cite{tomamichel2011}. Lower bounds on this quantity can be computed using a variety of methods under different assumptions. A simple approach, valid assuming independent and identically distributed (i.i.d.) behaviour of the devices in each round, is to compute the single-round conditional min-entropy. This can be formulated as a single semidefinite program (SDP), optimising over Eve's strategies. This approach was previously common in the literature. However, i.i.d. is a restrictive assumption which is often not easy to justify in implementations and at the same time, the bounds can be suboptimal \cite{carceller2025_improving}. A more powerful approach leverages the (generalised) entropy accumulation theorem \cite{ArnonFriedman2018,Dupuis2020,metger2022,Metger2023}, to lower bound the total conditional min-entropy in terms of the single-round conditional von Neumann entropy. This does not require an i.i.d.\ assumption and also enables accounting for finite-size effects. The von Neumann entropy is more difficult to compute but recent work has led to several numerical techniques based on SDP relaxations that are tractable for relevant device-independent and semi-device-independent scenarios \cite{Brown2024,carceller2025_improving, carceller2025_photon,kossmann2026}. Here, we implement such a method based on \cite{carceller2025_photon} and also compare it against simpler i.i.d.\ single-round min-entropy bounds.

In our model, the source is trusted while the measurement device is considered uncharacterised and may be correlated with Eve. It is associated with a $d$-outcome positive-operator-valued measure (POVM) $\{\hat{\Pi}_k^\lambda\}_k$, where $\lambda$ labels distinct measurement strategies that may vary between rounds. While the user only observes the averaged behaviour, Eve has access to $\lambda$. In \appref{app.shannon}, we derive an SDP relaxation lower bounding the conditional von Neumann entropy $H(B|E)$ of the output relative to Eve, based on the Gauss-Radau quadrature approximation to the relative entropy. The corresponding dual SDP takes the form
\begin{align}\label{eq:shannon_obj}
H(B|E) \!\geq\! c_m \!-\! \sum_{j=1}^{m-1}  \min_{\left\{\substack{
   \Gamma^{a}_{k},\ \nu_{bx} \\
   P,\ Q_{\omega}^{a}
}\right\}} \sum_{k,i}\nu_{ki}p(k|\rho_i)\! +\! \Tr\left[P\right] ,
\end{align}
where $c_m=\sum_{j=1}^{m}\tau_j$ and $\tau_j=\frac{\omega_j}{t_j\log 2}$, for $\omega_j$ and $t_j$ being the weights and nodes of the Gauss-Radau quadrature. The minimisation runs over all positive semidefinite matrices $\Gamma^{a}_{k}$ with a constrained block structure. Namely, the first block-element of $\Gamma^{a}_{k}$ is restricted by 
\begin{align}
    \sum_{a} \left(\Gamma_{k}^{a}\right)_{0,0} = \sum_{i}\nu_{ki}\rho_i + P \ .
\end{align}
Otherwise, if $u+v=1$ then,
\begin{align}
    \sum_{u,v} \left(\Gamma_{k}^{a}\right)_{u,v} = 2\tau_j\rho_{0}\delta_{a,k} + Q_{1}^{a}\! -\! \frac{1}{d}\Tr\left[Q_{1}^{a}\right]\mathbb{I} ,
\end{align}
if $v+u=2$ then,
\begin{align}
    \sum_{u,v} \left(\Gamma_{k}^{a}\right)_{u,v}\!\! = \tau_j\rho_{0}\!\left[\left(1\!-\!t_j\right)\delta_{a,k}\!+\!t_j\right]\! +\! Q_{2}^{a}\! -\! \frac{1}{d}\Tr\left[Q_{2}^{a}\right]\mathbb{I} ,
\end{align}
and finally if $u+v=s$ for $s >2$ then,
\begin{align}
    \sum_{u,v} \left(\Gamma_{k}^{a}\right)_{u,v} = Q_{s}^{a}\! -\! \frac{1}{d}\Tr\left[Q_{s}^{a}\right]\mathbb{I} ,
\end{align}
where $\nu_{ki}$ are dual scalar variables, and $P$ and $Q_{s}^{a}$ are dual $D\times D$ matrix variables, where $D$ is the dimension of the POVM elements. Note that while ideally we do not want to constrain the dimension, when implementing the SDP, $D$ must be finite. $D$ should thus be chosen sufficiently large to not affect the optimum. The quality of the approximation yielded by the SDP relaxation improves with growing the size of $\Gamma_{k}^{a}$, with $2D\times 2D$ being its minimum size for the SDP to be computable. For details, we refer to \appref{app.shannon}.

\section{Model}

We now apply our scheme to the particular case of homodyning the vacuum and probing with coherent states. We will model both additive Gaussian noise in the measurement and non-Gaussian noise from the analogue-to-digital converter (ADC), in order to compute how much randomness one may expect to extract from such a protocol. We stress, however, that in bounding the entropy, no assumptions are made about the particular form of the noise or the measurement.

The setup is illustrated in the lower panel of \figref{fig.concept}. The trusted input states are $\rho_0 = \ketbra{0}{0}$ and $\rho_i = \ketbra{\alpha_i}{\alpha_i}$, with $\alpha_i$ real, i.e.\ along the $X$-quadrature. We take the untrusted measurement to be of the $X$-quadrature, binned by a $\Delta$-bit ADC to give $d=2^\Delta$ outcomes. In the absence of imperfections, the measurement before binning is a projection $\ketbra{x}{x}$ onto quadrature eigenstates. To include noise, we consider a Gaussian distribution of variance $\noisevar$ and mean $x$ that converts the projector $\ketbra{x}{x}$ into a POVM element
\begin{equation}
\label{eq.noisyhomPOVM}
    \hat{\Sigma}_{x} = \int_{-\infty}^\infty \ket{y}\bra{y}\frac{\exp\left[-(y-x)^2/2\noisevar\right]}{\sqrt{2\pi} \noisestd}\,\dd y.
\end{equation}
The noise strength is thus determined by $\noisestd$ and can be stated in terms of the signal-to-noise ratio (SNR) $\vacvar/\noisevar$, where $\vacvar$ is the vacuum variance. The noise is additive as the observed variance for a Gaussian input state with variance $\vacvar$ becomes $\vacvar + \noisevar$. We also account for a detection efficiency $\eta$ (which transforms $\alpha_i \rightarrow \sqrt{\eta}\alpha_i$ in the calculation of the outcome distributions).

The ADC bins continuous outcomes into $d=2^\Delta$ intervals $I_k$, where $k$ labels the corresponding outputs. For an ideal ADC, the POVM elements thus become
\begin{equation}
\label{eq.binnedhomPOVM}
    \hat{\Sigma}_{k} = \int_{I_k}\hat{\Sigma}_{x}\, \dd x ,
\end{equation}
with $\hat{\Sigma}_{x}$ given by \eqref{eq.noisyhomPOVM}. Commonly, the $I_k$ are $d-2$ adjacent bins of equal width that span a range $R$ on both sides of a central point in the $X$-quadrature. The remaining two end bins cover the values below and above this range. Mathematically, we have
\begin{equation}
I_k=
\begin{cases}
    (-\infty, -R] & \text{if } k = 0\\
     [a_k-\delta/2, a_k+\delta /2) & \text{if } k=1,\dotsc, d-2\\
     (R, \infty) & \text{if } k=d-1
    \end{cases}
    \label{eq.fixedBinFunction}
\end{equation}
where $a_k=-R + (2k-1)\delta/2$ and $\delta=2R/(d-2)$ is the bin width. A priori, for randomness generation from the vacuum, a uniform distribution over $k$ is desired and the bin widths should then vary with $k$, since the corresponding $X$-quadrature distribution is a Gaussian. However, this is not always optimal here where additional probe states are also considered. Moreover, it is difficult to implement with standard ADCs, and thus less relevant experimentally. For completeness, in \appref{app.varbins} we do consider varying bin widths.

Modern ADCs achieve high speeds by interleaving several individual ADC units in time. This can lead to a non-Gaussian noise effect where the weights of adjacent even and odd bins become imbalanced. We model the corresponding POVM as
\begin{equation}
\label{eq.noisyADCpovm}
\hat{\Sigma}^\gamma_{k}=
\begin{cases}
    \hat{\Sigma}_{k} + \gamma \hat{\Sigma}_{k+1}\quad  &\text{($k$ even)} \\
     (1-\gamma)\hat{\Sigma}_{k} & \text{($k$ odd)}
    \end{cases}
\end{equation}
which can be understood as follows. For each odd bin, a fraction $\gamma$ of events corresponding to this outcome are shifted to the neighbouring even bin above. Given this model of the imperfect measurement device, we can compute the outcome distributions $p(k|\rho_i)$ needed in the minimisation \eqref{eq:shannon_obj}. We have
\begin{equation}
p^\gamma(k|\alpha)=
\begin{cases}
   p(k|\alpha) + \gamma p(k+1|\alpha)&\text{($k$ even)} \\
     (1-\gamma)p(k|\alpha) & \text{($k$ odd)}
    \end{cases}
\end{equation}
where,  as shown in \appref{app.cohprobs},
\begin{align}
    p(k|\alpha) &= \Tr[\ket{\alpha}\bra{\alpha}\hat{\Sigma}_{k}]\\
    &=\frac{1}{2}\erf\left[\upsilon(b_k)\right]-\frac{1}{2}\erf\left[\upsilon(a_k)\right],
\end{align}
where $a_k$ and $b_k$ are the lower and upper boundaries of $I_k$, respectively, and $\upsilon(x) = (x-\sqrt{2}\alpha)/(\sqrt{2}\totstd)$ with $\totstd$ the standard deviation of the total noise (vacuum and excess Gaussian noise).


\begin{figure}[t!]
    \includegraphics[width=\linewidth]{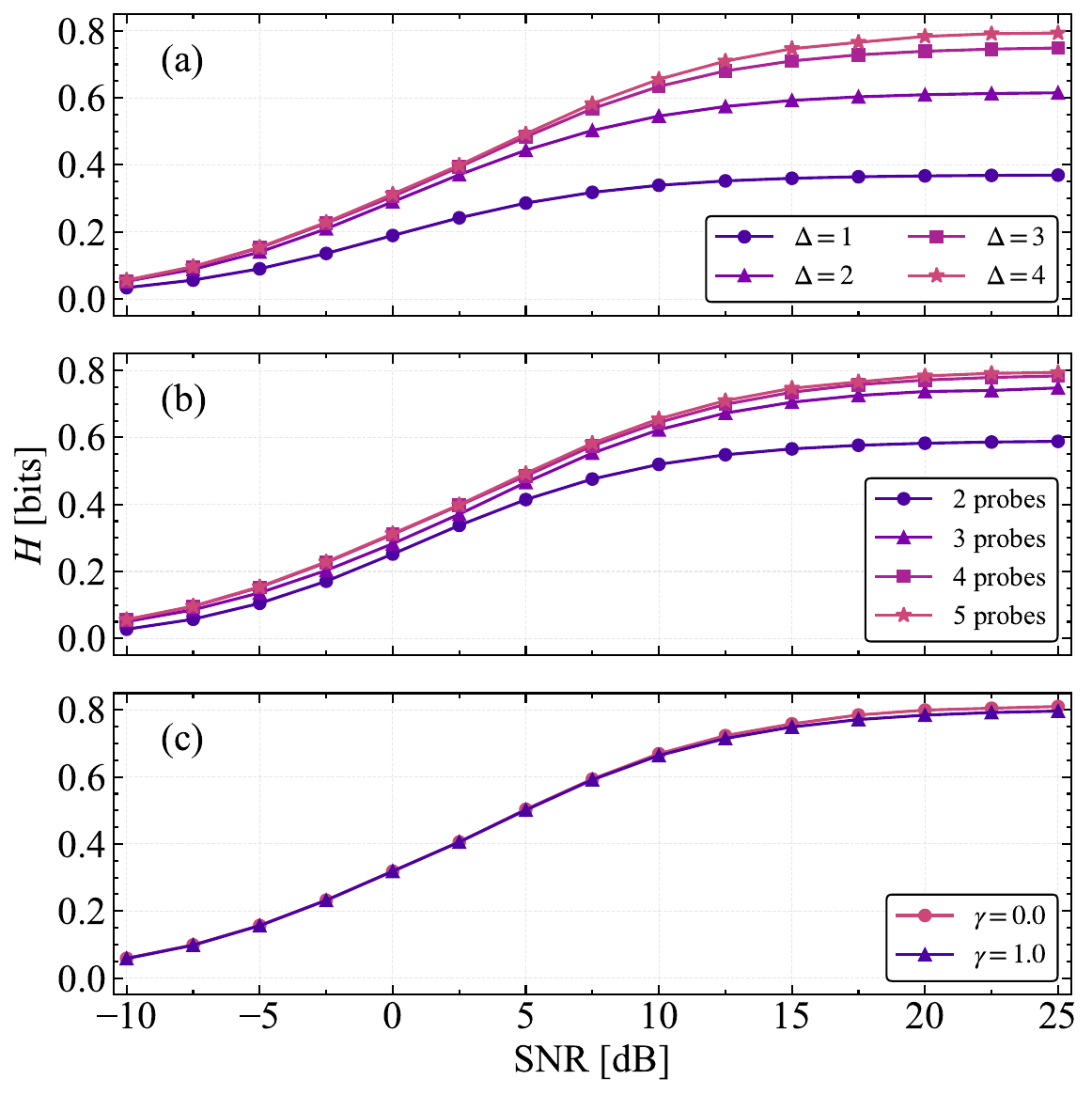}
	\caption{Results from modeling the protocol. The von Neumann entropy is shown vs.~the signal-to-noise ratio of the Gaussian noise, for varying parameter settings. \textbf{(a)} varying $\Delta = 1, 2, 3, 4$ (bottom to top) with fixed $\gamma=0.25$, $\eta=0.9$, 5 probe states with amplitudes equally spaced between 0 and a maximum value $\bar{\alpha}$ which, as well as $R$, is optimized at each SNR. \textbf{(b)} Fixed $\gamma=0.25$, $\eta=0.9$, $\Delta=4$ with optimal $R$ and $\bar{\alpha}$, and varying number of probe states 2,3,4,5 (bottom to top). \textbf{(c)} Fixed $\eta=0.9$, $\Delta=4$, 5 probe states, optimal $R$ and $\bar{\alpha}$, and varying $\gamma = 0, 1$ (top to bottom). For all the plots, we use $D = 10$.}\label{fig.theoryresults}
\end{figure}


\begin{figure*}[t!]
    \includegraphics[width=\linewidth]{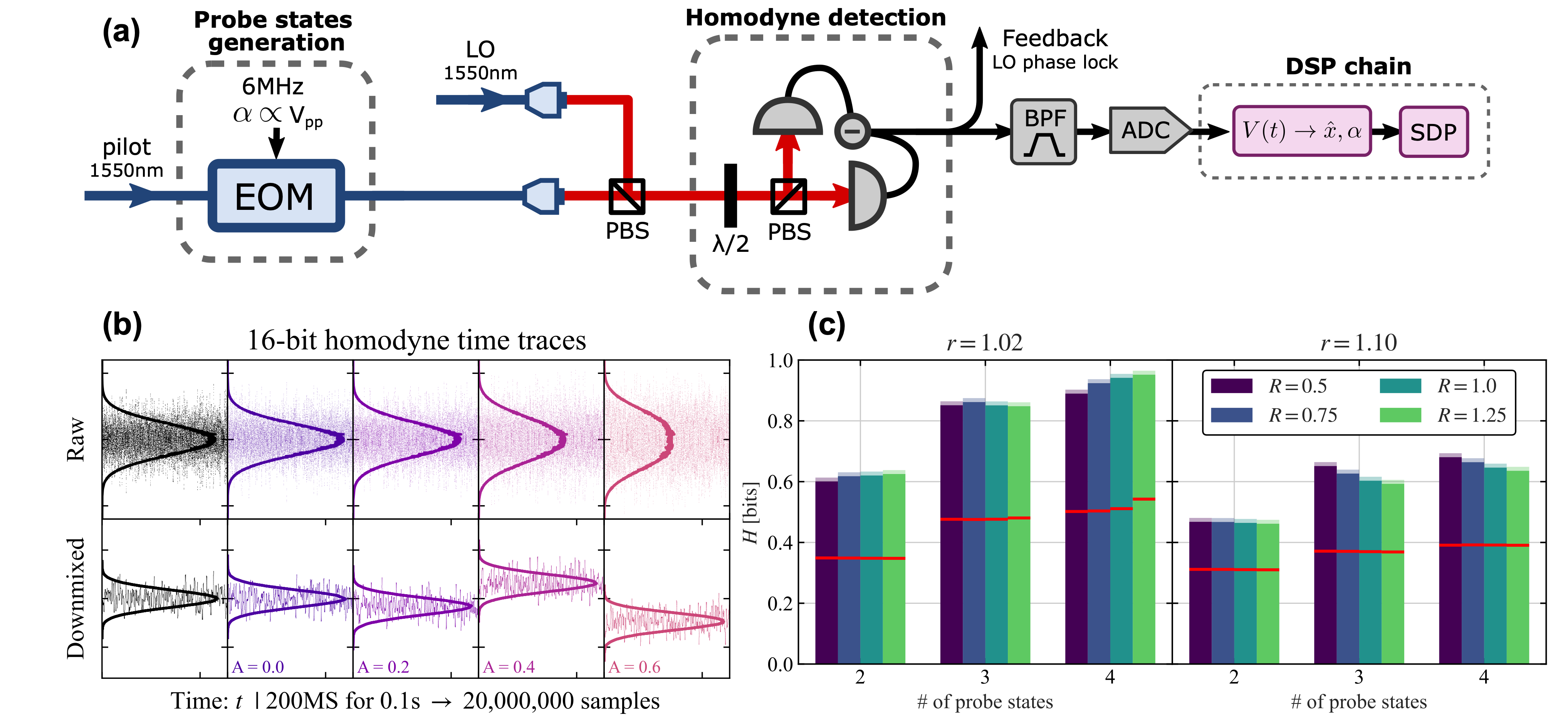}
	\caption{\textbf{(a)} Experimental setup. Probe states, created as sideband states by amplitude modulating a pilot beam with an EOM, are measured by balanced homodyne detection. The photocurrent is bandpass filtered (BPF) around the modulation frequency before being converted to bit values by an ADC. The resulting time traces are then processed by a 2-block digital signal processing (DSP) chain. \textbf{(b)} Time traces collected for a single data set overlain with equally-scaled histograms of the bit value distributions. For the raw time traces, the vertical axis spans the entire ADC range: [$-2^{15},2^{15}$], while for the downmixed time traces we zoom to the middle half. \textbf{(c)} Experimental von Neumann entropy for various DSP settings. We perform downsampling with bit depth $\Delta=4$ and various ADC ranges $R$. The faded histograms show the von Neumann entropy in the ideal case of infinite data, while the overlaid solid histograms show the entropy after accounting for finite-size effects. Horizontal red lines show single-round min-entropy. To handle uncertainty in the estimation of the probe state amplitudes, we consider two values of the scaling parameter $r$ equivalent to assuming the amplitudes are $2\%$ (left) and $10\%$ (right) larger than the estimated values (see \appref{sec:app_dsp3}).}
	\label{fig:expsetup}
\end{figure*}

The results for randomness certification are shown in \figref{fig.theoryresults}. We see that while the randomness degrades with very strong additive noise (low SNR), for reasonable noise at the 10\% level, the protocol performs well and the entropy starts to saturate. From \figref{fig.theoryresults}(a), we see that, as may be expected, more randomness can be certified with a larger number of outputs but saturates with increasing $\Delta$. For the parameter settings considered here, the additional gain above $\Delta=4$ (16 outputs) is minor \footnote{We note that, in the absence of noise and for a tomographically complete set of probe states (or equivalently a fully characterised POVM), the certifiable randomness will be $\Delta$ bits. However, the coherent-state probes considered here are not tomographically complete and even for a complete set, convergence towards maximal randomness with SNR is slow }. From \figref{fig.theoryresults}(b), we see that a similar behaviour applies to the number of trusted states. Notably, already with two states (i.e.~the vacuum and a single coherent state), randomness can be certified with $H \approx 0.6$ in a low-noise scenario. Finally, from  \figref{fig.theoryresults}(c) we see that the protocol is very robust against the non-Gaussian ADC noise. Even for $\gamma=1$, where every second bin is effectively removed, the entropy barely degrades in the optimal regime.

\section{Experiment}

Next, we experimentally demonstrate the performance of the protocol using the setup outlined in Fig. \ref{fig:expsetup}(a) (for a more detailed figure and explanation see App. \ref{sec:app_exp}). Probe states are generated as sideband states by intensity modulating a pilot beam at $f_\text{mod}=\SI{6}{\MHz}$ with an electro-optic modulator (EOM). The peak-to-peak voltage $V_{pp}$ used to drive the EOM is well below the corresponding half-wave voltage of the EOM, thus the amplitudes $\alpha$ of the sideband states are directly proportional to $V_{pp}$. In a single data set we collect five time traces: a shot-noise trace (corresponding to vacuum), where the pilot beam is blocked, and four traces with $A=0.0$, $0.2$, $0.4$, $0.6$, where $A$ indicates the scaling of $V_{pp}$. The pilot beam is then directly measured by a polarization-based homodyne detector and the resulting photo-current is bandpass filtered (BPF) around $f_\text{mod}$ before being recorded by a 16-bit ADC. The homodyne detector also has a lower-bandwidth DC output, which is used to generate an error signal to lock the phase between the local oscillator (LO) and pilot beam, such that the $X$-quadrature is measured. The bit value distributions $V(t)$, as recorded by the ADC of five such traces, are shown in the top row of Fig. \ref{fig:expsetup}(b). In order to optimise the certifiable randomness, the LO power and ADC range were tuned such that the shot noise trace spans the entire ADC range. The traces were recorded with an estimated total detection efficiency of at least $90\%$, with inefficiency stemming from propagation loss between the EOM and homodyne detector, imperfect visibility between pilot and LO and the quantum efficiency of the homodyne detector diodes.

After the ADC, the data set goes through a digital signal processing (DSP) chain consisting of two blocks. In the first block, the $X$-quadrature distributions of the sideband probe states are extracted from the raw time traces and their corresponding amplitudes estimated. This is done in four steps: digital downmixing, shot noise normalization, amplitude estimation and downsampling, which are all described in detail in \appref{sec:app_dsp}. The result of the first of these steps $V_\text{dm}(t)$ is shown in the bottom row of \figref{fig:expsetup}(b). In the second block, the observed distributions and estimated amplitudes are used to compute a bound on the entropy of the measured data from the vacuum input via SDP, as described above. Specifically, the trusted states in the SDP constraints are coherent states based on the estimated amplitudes and the observed distribution is given by the normalized histograms from the experiment. The result of this is shown in \figref{fig:expsetup}(c). In order to avoid estimation errors compromising the security, we scale our estimations of the amplitudes according to $\alpha_i=r\alpha^\text{true}_i$ with $r>1$, thus making the trusted states more distinguishable and decreasing the certifiable randomness (see \appref{app.amplitudeerr}). Additionally, we account for finite-size effects on the output entropy via the generalised entropy accumulation theorem \cite{metger2022} (see \appref{sec:finite_size}). 

The results presented in \figref{fig:expsetup}(c) show that a considerable amount of randomness can be generated using our protocol with a relatively small number of probe states after accounting for both amplitude estimation errors and finite-size effects. We observe that, in the case of low estimation errors and ignoring finite-size effects, the minimum entropy increases with $R$. However, the finite-size correction also increases with $R$, eliminating the advantage in most cases. Nevertheless, even with 10\% estimation errors the entropy per round remains above 0.4. The horizontal red lines in \figref{fig:expsetup}(c) show corresponding bounds based on the single-round min-entropy without any finite-size correction (see \appref{app.dual}). We note that these significantly underestimate the certifiable randomness even though they rely on the more restrictive i.i.d. assumption.

\section{Conclusion and outlook}

In summary, we have demonstrated a protocol for semi-device-independent randomness certification that is simple to implement yet highly robust and secure. The entropy can be bounded online (interleaving generation and additional probe rounds), allowing continuous self-testing operation. In future work, it would be interesting to allow for probe states correlated with Eve (mixed states) and to explore potential extensions to quantum key distribution or other protocols.

For the sampling rate of our implementation of 200 MS/s, the extractable secret bit rate is in the range of about 100-180 Mbps, depending on the choice of parameters (cf.\ \figref{fig:expsetup}(c)). This exceeds (but is comparable to) the highest reported rates for other schemes with untrusted measurement devices, e.g. \cite{Rusca2019,Avesani2021,Nie2024,Argillander2025,Sabatini2026}. Our entropy per round of about 0.5-0.9 bits is also, to our knowledge, the highest reported for untrusted continuous-variable measurements. However, some schemes with different semi-device-independent assumptions, in particular source-independent schemes with trusted measurements, have reached much higher rates, in the tens of Gbps, e.g.\ \cite{Avesani2018,Bian2025,Bertapelle2025,Cizauskas2025,Kincaid2025}. It is natural to ask how the rate of the scheme presented here might be improved. As can be seen from \figref{fig.theoryresults}, the entropy begins to saturate with increasing bit depth and number of probe states, hence increasing the entropy per round much beyond the one achieved here is probably not a viable route. The sampling rate, on the other hand, could potentially be significantly enhanced with fast detectors and electronics, allowing for Gbps secret bit rates.

Note that an experiment demonstrating a very similar protocol was developed in parallel by Wang et al.\ in the group of Charles Lim at the National University of Singapore, appearing online a few months before the first version of the present manuscript \cite{Wang2023}. In that work, a slightly different configuration of coherent states was used (varying both phase and amplitude). Approximately 0.005 of entropy per round could be certified, significantly lower than in the present work, but full randomness extraction was implemented via Toeplitz hashing.


\begin{acknowledgements}
We gratefully acknowledge support from the Danish National Research Foundation, Center for Macroscopic Quantum States (bigQ, DNRF142), the Carlsberg Foundation CF19-0313 and CF21-0466, the Independent Research Fund Denmark 7027-00044B and 0171-00055B,  the Innovation Fund Denmark 3200-00012B, JST SPRING JPMJSP2123, JST Moonshot R\&D JPMJMS226C, and JST ASPIRE JPMJAP2427. C.R.C. is supported by the Swedish Foundation for Strategic Research and the European Union’s Horizon 2020 research and innovation programme under the Marie Skłodowska-Curie grant agreement No 101262877, Project TPSQCrypto. Views and opinions expressed are however those of the author(s) only and do not necessarily reflect those of the European Union or European Research Executive Agency. Neither the European Union nor the granting authority can be held responsible for them.
\end{acknowledgements}

\bibliography{cohstate-SDI-QRNG_bibliography.bib}

\begin{thebibliography}{61}%
\makeatletter
\providecommand \@ifxundefined [1]{%
 \@ifx{#1\undefined}
}%
\providecommand \@ifnum [1]{%
 \ifnum #1\expandafter \@firstoftwo
 \else \expandafter \@secondoftwo
 \fi
}%
\providecommand \@ifx [1]{%
 \ifx #1\expandafter \@firstoftwo
 \else \expandafter \@secondoftwo
 \fi
}%
\providecommand \natexlab [1]{#1}%
\providecommand \enquote  [1]{``#1''}%
\providecommand \bibnamefont  [1]{#1}%
\providecommand \bibfnamefont [1]{#1}%
\providecommand \citenamefont [1]{#1}%
\providecommand \href@noop [0]{\@secondoftwo}%
\providecommand \href [0]{\begingroup \@sanitize@url \@href}%
\providecommand \@href[1]{\@@startlink{#1}\@@href}%
\providecommand \@@href[1]{\endgroup#1\@@endlink}%
\providecommand \@sanitize@url [0]{\catcode `\\12\catcode `\$12\catcode `\&12\catcode `\#12\catcode `\^12\catcode `\_12\catcode `\%12\relax}%
\providecommand \@@startlink[1]{}%
\providecommand \@@endlink[0]{}%
\providecommand \url  [0]{\begingroup\@sanitize@url \@url }%
\providecommand \@url [1]{\endgroup\@href {#1}{\urlprefix }}%
\providecommand \urlprefix  [0]{URL }%
\providecommand \Eprint [0]{\href }%
\providecommand \doibase [0]{https://doi.org/}%
\providecommand \selectlanguage [0]{\@gobble}%
\providecommand \bibinfo  [0]{\@secondoftwo}%
\providecommand \bibfield  [0]{\@secondoftwo}%
\providecommand \translation [1]{[#1]}%
\providecommand \BibitemOpen [0]{}%
\providecommand \bibitemStop [0]{}%
\providecommand \bibitemNoStop [0]{.\EOS\space}%
\providecommand \EOS [0]{\spacefactor3000\relax}%
\providecommand \BibitemShut  [1]{\csname bibitem#1\endcsname}%
\let\auto@bib@innerbib\@empty
\bibitem [{\citenamefont {Hayes}(2001)}]{Hayes2001}%
  \BibitemOpen
  \bibfield  {author} {\bibinfo {author} {\bibfnamefont {B.}~\bibnamefont {Hayes}},\ }\bibfield  {title} {\bibinfo {title} {Randomness as a resource},\ }\href {https://doi.org/10.1511/2001.28.3336} {\bibfield  {journal} {\bibinfo  {journal} {American Scientist}\ }\textbf {\bibinfo {volume} {89}},\ \bibinfo {pages} {300} (\bibinfo {year} {2001})}\BibitemShut {NoStop}%
\bibitem [{\citenamefont {Ac\'in}\ and\ \citenamefont {Masanes}(2016)}]{Acin2016}%
  \BibitemOpen
  \bibfield  {author} {\bibinfo {author} {\bibfnamefont {A.}~\bibnamefont {Ac\'in}}\ and\ \bibinfo {author} {\bibfnamefont {L.}~\bibnamefont {Masanes}},\ }\bibfield  {title} {\bibinfo {title} {Certified randomness in quantum physics},\ }\href {https://doi.org/doi:10.1038/nature20119} {\bibfield  {journal} {\bibinfo  {journal} {Nature}\ }\textbf {\bibinfo {volume} {540}},\ \bibinfo {pages} {213} (\bibinfo {year} {2016})}\BibitemShut {NoStop}%
\bibitem [{\citenamefont {Herrero-Collantes}\ and\ \citenamefont {Garcia-Escartin}(2017)}]{Herrero2017}%
  \BibitemOpen
  \bibfield  {author} {\bibinfo {author} {\bibfnamefont {M.}~\bibnamefont {Herrero-Collantes}}\ and\ \bibinfo {author} {\bibfnamefont {J.~C.}\ \bibnamefont {Garcia-Escartin}},\ }\bibfield  {title} {\bibinfo {title} {Quantum random number generators},\ }\href {https://doi.org/10.1103/RevModPhys.89.015004} {\bibfield  {journal} {\bibinfo  {journal} {Rev. Mod. Phys.}\ }\textbf {\bibinfo {volume} {89}},\ \bibinfo {pages} {015004} (\bibinfo {year} {2017})}\BibitemShut {NoStop}%
\bibitem [{\citenamefont {Bera}\ \emph {et~al.}(2017)\citenamefont {Bera}, \citenamefont {Ac\'in}, \citenamefont {Kus}, \citenamefont {Mitchell},\ and\ \citenamefont {Lewenstein}}]{Bera2017}%
  \BibitemOpen
  \bibfield  {author} {\bibinfo {author} {\bibfnamefont {M.}~\bibnamefont {Bera}}, \bibinfo {author} {\bibfnamefont {A.}~\bibnamefont {Ac\'in}}, \bibinfo {author} {\bibfnamefont {M.}~\bibnamefont {Kus}}, \bibinfo {author} {\bibfnamefont {M.}~\bibnamefont {Mitchell}},\ and\ \bibinfo {author} {\bibfnamefont {M.}~\bibnamefont {Lewenstein}},\ }\bibfield  {title} {\bibinfo {title} {{Randomness in Quantum Mechanics: Philosophy, Physics and Technology}},\ }\href {https://doi.org/10.1088/1361-6633/aa8731} {\bibfield  {journal} {\bibinfo  {journal} {Rep. Prog. Phys.}\ }\textbf {\bibinfo {volume} {80}},\ \bibinfo {pages} {124001} (\bibinfo {year} {2017})}\BibitemShut {NoStop}%
\bibitem [{\citenamefont {Stefanov}\ \emph {et~al.}(2000)\citenamefont {Stefanov}, \citenamefont {Gisin}, \citenamefont {Guinnard}, \citenamefont {Guinnard},\ and\ \citenamefont {Zbinden}}]{Stefanov2000}%
  \BibitemOpen
  \bibfield  {author} {\bibinfo {author} {\bibfnamefont {A.}~\bibnamefont {Stefanov}}, \bibinfo {author} {\bibfnamefont {N.}~\bibnamefont {Gisin}}, \bibinfo {author} {\bibfnamefont {O.}~\bibnamefont {Guinnard}}, \bibinfo {author} {\bibfnamefont {L.}~\bibnamefont {Guinnard}},\ and\ \bibinfo {author} {\bibfnamefont {H.}~\bibnamefont {Zbinden}},\ }\bibfield  {title} {\bibinfo {title} {Optical quantum random number generator},\ }\href {https://doi.org/10.1080/09500340008233380} {\bibfield  {journal} {\bibinfo  {journal} {J. Mod. Opt.}\ }\textbf {\bibinfo {volume} {47}},\ \bibinfo {pages} {595} (\bibinfo {year} {2000})}\BibitemShut {NoStop}%
\bibitem [{\citenamefont {Bell}(1964)}]{Bell1964}%
  \BibitemOpen
  \bibfield  {author} {\bibinfo {author} {\bibfnamefont {J.}~\bibnamefont {Bell}},\ }\bibfield  {title} {\bibinfo {title} {On the einstein podolsky rosen paradox},\ }\href {https://doi.org/10.1103/PhysicsPhysiqueFizika.1.195} {\bibfield  {journal} {\bibinfo  {journal} {Physics}\ }\textbf {\bibinfo {volume} {1}},\ \bibinfo {pages} {195} (\bibinfo {year} {1964})}\BibitemShut {NoStop}%
\bibitem [{\citenamefont {Brunner}\ \emph {et~al.}(2014)\citenamefont {Brunner}, \citenamefont {Cavalcanti}, \citenamefont {Pironio}, \citenamefont {Scarani},\ and\ \citenamefont {Wehner}}]{Brunner2014}%
  \BibitemOpen
  \bibfield  {author} {\bibinfo {author} {\bibfnamefont {N.}~\bibnamefont {Brunner}}, \bibinfo {author} {\bibfnamefont {D.}~\bibnamefont {Cavalcanti}}, \bibinfo {author} {\bibfnamefont {S.}~\bibnamefont {Pironio}}, \bibinfo {author} {\bibfnamefont {V.}~\bibnamefont {Scarani}},\ and\ \bibinfo {author} {\bibfnamefont {S.}~\bibnamefont {Wehner}},\ }\bibfield  {title} {\bibinfo {title} {Bell nonlocality},\ }\href {https://doi.org/10.1103/RevModPhys.86.419} {\bibfield  {journal} {\bibinfo  {journal} {Rev. Mod. Phys.}\ }\textbf {\bibinfo {volume} {86}},\ \bibinfo {pages} {419} (\bibinfo {year} {2014})}\BibitemShut {NoStop}%
\bibitem [{\citenamefont {Colbeck}(2009)}]{colbeckPhD2009}%
  \BibitemOpen
  \bibfield  {author} {\bibinfo {author} {\bibfnamefont {R.}~\bibnamefont {Colbeck}},\ }\href {https://arxiv.org/abs/0911.3814} {\bibinfo {title} {Quantum and relativistic protocols for secure multi-party computation}},\ \bibinfo {howpublished} {Ph.D. Thesis, University of Cambridge} (\bibinfo {year} {2009}),\ \bibinfo {note} {arXiv:0911.3814 [quant-ph]}\BibitemShut {NoStop}%
\bibitem [{\citenamefont {Pironio}\ \emph {et~al.}(2010)\citenamefont {Pironio}, \citenamefont {Ac\'in}, \citenamefont {Massar}, \citenamefont {de~la Giroday}, \citenamefont {Matsukevich}, \citenamefont {Maunz}, \citenamefont {Olmschenk}, \citenamefont {Hayes}, \citenamefont {Luo}, \citenamefont {Manning},\ and\ \citenamefont {Monroe}}]{Pironio2010}%
  \BibitemOpen
  \bibfield  {author} {\bibinfo {author} {\bibfnamefont {S.}~\bibnamefont {Pironio}}, \bibinfo {author} {\bibfnamefont {A.}~\bibnamefont {Ac\'in}}, \bibinfo {author} {\bibfnamefont {S.}~\bibnamefont {Massar}}, \bibinfo {author} {\bibfnamefont {A.~B.}\ \bibnamefont {de~la Giroday}}, \bibinfo {author} {\bibfnamefont {D.~N.}\ \bibnamefont {Matsukevich}}, \bibinfo {author} {\bibfnamefont {P.}~\bibnamefont {Maunz}}, \bibinfo {author} {\bibfnamefont {S.}~\bibnamefont {Olmschenk}}, \bibinfo {author} {\bibfnamefont {D.}~\bibnamefont {Hayes}}, \bibinfo {author} {\bibfnamefont {L.}~\bibnamefont {Luo}}, \bibinfo {author} {\bibfnamefont {T.~A.}\ \bibnamefont {Manning}},\ and\ \bibinfo {author} {\bibfnamefont {C.}~\bibnamefont {Monroe}},\ }\bibfield  {title} {\bibinfo {title} {Random numbers certified by {Bell's} theorem},\ }\href {https://doi.org/doi:10.1038/nature09008} {\bibfield  {journal} {\bibinfo  {journal} {Nature}\ }\textbf {\bibinfo {volume} {464}},\ \bibinfo {pages} {1021} (\bibinfo {year} {2010})}\BibitemShut
  {NoStop}%
\bibitem [{\citenamefont {Christensen}\ \emph {et~al.}(2013)\citenamefont {Christensen}, \citenamefont {McCusker}, \citenamefont {Altepeter}, \citenamefont {Calkins}, \citenamefont {Gerrits}, \citenamefont {Lita}, \citenamefont {Miller}, \citenamefont {Shalm}, \citenamefont {Zhang}, \citenamefont {Nam}, \citenamefont {Brunner}, \citenamefont {Lim}, \citenamefont {Gisin},\ and\ \citenamefont {Kwiat}}]{Christensen2013}%
  \BibitemOpen
  \bibfield  {author} {\bibinfo {author} {\bibfnamefont {B.~G.}\ \bibnamefont {Christensen}}, \bibinfo {author} {\bibfnamefont {K.~T.}\ \bibnamefont {McCusker}}, \bibinfo {author} {\bibfnamefont {J.~B.}\ \bibnamefont {Altepeter}}, \bibinfo {author} {\bibfnamefont {B.}~\bibnamefont {Calkins}}, \bibinfo {author} {\bibfnamefont {T.}~\bibnamefont {Gerrits}}, \bibinfo {author} {\bibfnamefont {A.~E.}\ \bibnamefont {Lita}}, \bibinfo {author} {\bibfnamefont {A.}~\bibnamefont {Miller}}, \bibinfo {author} {\bibfnamefont {L.~K.}\ \bibnamefont {Shalm}}, \bibinfo {author} {\bibfnamefont {Y.}~\bibnamefont {Zhang}}, \bibinfo {author} {\bibfnamefont {S.~W.}\ \bibnamefont {Nam}}, \bibinfo {author} {\bibfnamefont {N.}~\bibnamefont {Brunner}}, \bibinfo {author} {\bibfnamefont {C.~C.~W.}\ \bibnamefont {Lim}}, \bibinfo {author} {\bibfnamefont {N.}~\bibnamefont {Gisin}},\ and\ \bibinfo {author} {\bibfnamefont {P.~G.}\ \bibnamefont {Kwiat}},\ }\bibfield  {title} {\bibinfo {title} {Detection-loophole-free test of quantum
  nonlocality, and applications},\ }\href {https://doi.org/10.1103/PhysRevLett.111.130406} {\bibfield  {journal} {\bibinfo  {journal} {Phys. Rev. Lett.}\ }\textbf {\bibinfo {volume} {111}},\ \bibinfo {pages} {130406} (\bibinfo {year} {2013})}\BibitemShut {NoStop}%
\bibitem [{\citenamefont {Bierhorst}\ \emph {et~al.}(2018)\citenamefont {Bierhorst}, \citenamefont {Knill}, \citenamefont {Glancy}, \citenamefont {Zhang}, \citenamefont {Mink}, \citenamefont {Jordan}, \citenamefont {Rommal}, \citenamefont {Liu}, \citenamefont {Christensen}, \citenamefont {Nam}, \citenamefont {Stevens},\ and\ \citenamefont {Shalm}}]{Bierhorst2018}%
  \BibitemOpen
  \bibfield  {author} {\bibinfo {author} {\bibfnamefont {P.}~\bibnamefont {Bierhorst}}, \bibinfo {author} {\bibfnamefont {E.}~\bibnamefont {Knill}}, \bibinfo {author} {\bibfnamefont {S.}~\bibnamefont {Glancy}}, \bibinfo {author} {\bibfnamefont {Y.}~\bibnamefont {Zhang}}, \bibinfo {author} {\bibfnamefont {A.}~\bibnamefont {Mink}}, \bibinfo {author} {\bibfnamefont {S.}~\bibnamefont {Jordan}}, \bibinfo {author} {\bibfnamefont {A.}~\bibnamefont {Rommal}}, \bibinfo {author} {\bibfnamefont {Y.-K.}\ \bibnamefont {Liu}}, \bibinfo {author} {\bibfnamefont {B.}~\bibnamefont {Christensen}}, \bibinfo {author} {\bibfnamefont {S.~W.}\ \bibnamefont {Nam}}, \bibinfo {author} {\bibfnamefont {M.~J.}\ \bibnamefont {Stevens}},\ and\ \bibinfo {author} {\bibfnamefont {L.~K.}\ \bibnamefont {Shalm}},\ }\bibfield  {title} {\bibinfo {title} {Experimentally {Generated} {Randomness} {Certified} by the {Impossibility} of {Superluminal} {Signals}},\ }\href {https://doi.org/10.1038/s41586-018-0019-0} {\bibfield  {journal} {\bibinfo
  {journal} {Nature}\ }\textbf {\bibinfo {volume} {556}},\ \bibinfo {pages} {223} (\bibinfo {year} {2018})}\BibitemShut {NoStop}%
\bibitem [{\citenamefont {Liu}\ \emph {et~al.}(2018)\citenamefont {Liu}, \citenamefont {Zhao}, \citenamefont {Li}, \citenamefont {Guan}, \citenamefont {Zhang}, \citenamefont {Bai}, \citenamefont {Zhang}, \citenamefont {Liu}, \citenamefont {Wu}, \citenamefont {Yuan}, \citenamefont {Li}, \citenamefont {Munro}, \citenamefont {Wang}, \citenamefont {You}, \citenamefont {Zhang}, \citenamefont {Ma}, \citenamefont {Fan}, \citenamefont {Zhang},\ and\ \citenamefont {Pan}}]{Liu2018}%
  \BibitemOpen
  \bibfield  {author} {\bibinfo {author} {\bibfnamefont {Y.}~\bibnamefont {Liu}}, \bibinfo {author} {\bibfnamefont {Q.}~\bibnamefont {Zhao}}, \bibinfo {author} {\bibfnamefont {M.-H.}\ \bibnamefont {Li}}, \bibinfo {author} {\bibfnamefont {J.-Y.}\ \bibnamefont {Guan}}, \bibinfo {author} {\bibfnamefont {Y.}~\bibnamefont {Zhang}}, \bibinfo {author} {\bibfnamefont {B.}~\bibnamefont {Bai}}, \bibinfo {author} {\bibfnamefont {W.}~\bibnamefont {Zhang}}, \bibinfo {author} {\bibfnamefont {W.-Z.}\ \bibnamefont {Liu}}, \bibinfo {author} {\bibfnamefont {C.}~\bibnamefont {Wu}}, \bibinfo {author} {\bibfnamefont {X.}~\bibnamefont {Yuan}}, \bibinfo {author} {\bibfnamefont {H.}~\bibnamefont {Li}}, \bibinfo {author} {\bibfnamefont {W.~J.}\ \bibnamefont {Munro}}, \bibinfo {author} {\bibfnamefont {Z.}~\bibnamefont {Wang}}, \bibinfo {author} {\bibfnamefont {L.}~\bibnamefont {You}}, \bibinfo {author} {\bibfnamefont {J.}~\bibnamefont {Zhang}}, \bibinfo {author} {\bibfnamefont {X.}~\bibnamefont {Ma}}, \bibinfo {author} {\bibfnamefont
  {J.}~\bibnamefont {Fan}}, \bibinfo {author} {\bibfnamefont {Q.}~\bibnamefont {Zhang}},\ and\ \bibinfo {author} {\bibfnamefont {J.-W.}\ \bibnamefont {Pan}},\ }\bibfield  {title} {\bibinfo {title} {Device-independent quantum random-number generation},\ }\href {https://doi.org/10.1038/s41586-018-0559-3} {\bibfield  {journal} {\bibinfo  {journal} {Nature}\ }\textbf {\bibinfo {volume} {562}},\ \bibinfo {pages} {548} (\bibinfo {year} {2018})}\BibitemShut {NoStop}%
\bibitem [{\citenamefont {Shalm}\ \emph {et~al.}(2021)\citenamefont {Shalm}, \citenamefont {Zhang}, \citenamefont {Bienfang}, \citenamefont {Schlager}, \citenamefont {Stevens}, \citenamefont {Mazurek}, \citenamefont {Abellán}, \citenamefont {Amaya}, \citenamefont {Mitchell}, \citenamefont {Alhejji}, \citenamefont {Fu}, \citenamefont {Ornstein}, \citenamefont {Mirin}, \citenamefont {Nam},\ and\ \citenamefont {Knill}}]{Shalm2021}%
  \BibitemOpen
  \bibfield  {author} {\bibinfo {author} {\bibfnamefont {L.~K.}\ \bibnamefont {Shalm}}, \bibinfo {author} {\bibfnamefont {Y.}~\bibnamefont {Zhang}}, \bibinfo {author} {\bibfnamefont {J.~C.}\ \bibnamefont {Bienfang}}, \bibinfo {author} {\bibfnamefont {C.}~\bibnamefont {Schlager}}, \bibinfo {author} {\bibfnamefont {M.~J.}\ \bibnamefont {Stevens}}, \bibinfo {author} {\bibfnamefont {M.~D.}\ \bibnamefont {Mazurek}}, \bibinfo {author} {\bibfnamefont {C.}~\bibnamefont {Abellán}}, \bibinfo {author} {\bibfnamefont {W.}~\bibnamefont {Amaya}}, \bibinfo {author} {\bibfnamefont {M.~W.}\ \bibnamefont {Mitchell}}, \bibinfo {author} {\bibfnamefont {M.~A.}\ \bibnamefont {Alhejji}}, \bibinfo {author} {\bibfnamefont {H.}~\bibnamefont {Fu}}, \bibinfo {author} {\bibfnamefont {J.}~\bibnamefont {Ornstein}}, \bibinfo {author} {\bibfnamefont {R.~P.}\ \bibnamefont {Mirin}}, \bibinfo {author} {\bibfnamefont {S.~W.}\ \bibnamefont {Nam}},\ and\ \bibinfo {author} {\bibfnamefont {E.}~\bibnamefont {Knill}},\ }\bibfield  {title} {\bibinfo
  {title} {Device-independent randomness expansion with entangled photons},\ }\href {https://doi.org/10.1038/s41567-020-01153-4} {\bibfield  {journal} {\bibinfo  {journal} {Nature Physics}\ }\textbf {\bibinfo {volume} {17}},\ \bibinfo {pages} {452} (\bibinfo {year} {2021})}\BibitemShut {NoStop}%
\bibitem [{\citenamefont {Liu}\ \emph {et~al.}(2021)\citenamefont {Liu}, \citenamefont {Li}, \citenamefont {Ragy}, \citenamefont {Zhao}, \citenamefont {Bai}, \citenamefont {Liu}, \citenamefont {Brown}, \citenamefont {Zhang}, \citenamefont {Colbeck}, \citenamefont {Fan}, \citenamefont {Zhang},\ and\ \citenamefont {Pan}}]{Liu2021}%
  \BibitemOpen
  \bibfield  {author} {\bibinfo {author} {\bibfnamefont {W.-Z.}\ \bibnamefont {Liu}}, \bibinfo {author} {\bibfnamefont {M.-H.}\ \bibnamefont {Li}}, \bibinfo {author} {\bibfnamefont {S.}~\bibnamefont {Ragy}}, \bibinfo {author} {\bibfnamefont {S.-R.}\ \bibnamefont {Zhao}}, \bibinfo {author} {\bibfnamefont {B.}~\bibnamefont {Bai}}, \bibinfo {author} {\bibfnamefont {Y.}~\bibnamefont {Liu}}, \bibinfo {author} {\bibfnamefont {P.~J.}\ \bibnamefont {Brown}}, \bibinfo {author} {\bibfnamefont {J.}~\bibnamefont {Zhang}}, \bibinfo {author} {\bibfnamefont {R.}~\bibnamefont {Colbeck}}, \bibinfo {author} {\bibfnamefont {J.}~\bibnamefont {Fan}}, \bibinfo {author} {\bibfnamefont {Q.}~\bibnamefont {Zhang}},\ and\ \bibinfo {author} {\bibfnamefont {J.-W.}\ \bibnamefont {Pan}},\ }\bibfield  {title} {\bibinfo {title} {Device-independent randomness expansion against quantum side information},\ }\href {https://doi.org/10.1038/s41567-020-01147-2} {\bibfield  {journal} {\bibinfo  {journal} {Nature Physics}\ }\textbf {\bibinfo {volume}
  {17}},\ \bibinfo {pages} {448} (\bibinfo {year} {2021})}\BibitemShut {NoStop}%
\bibitem [{\citenamefont {Li}\ \emph {et~al.}(2011)\citenamefont {Li}, \citenamefont {Yin}, \citenamefont {Wu}, \citenamefont {Zou}, \citenamefont {Wang}, \citenamefont {Chen}, \citenamefont {Guo},\ and\ \citenamefont {Han}}]{Li2011}%
  \BibitemOpen
  \bibfield  {author} {\bibinfo {author} {\bibfnamefont {H.-W.}\ \bibnamefont {Li}}, \bibinfo {author} {\bibfnamefont {Z.-Q.}\ \bibnamefont {Yin}}, \bibinfo {author} {\bibfnamefont {Y.-C.}\ \bibnamefont {Wu}}, \bibinfo {author} {\bibfnamefont {X.-B.}\ \bibnamefont {Zou}}, \bibinfo {author} {\bibfnamefont {S.}~\bibnamefont {Wang}}, \bibinfo {author} {\bibfnamefont {W.}~\bibnamefont {Chen}}, \bibinfo {author} {\bibfnamefont {G.-C.}\ \bibnamefont {Guo}},\ and\ \bibinfo {author} {\bibfnamefont {Z.-F.}\ \bibnamefont {Han}},\ }\bibfield  {title} {\bibinfo {title} {Semi-device-independent random-number expansion without entanglement},\ }\href {https://doi.org/10.1103/PhysRevA.84.034301} {\bibfield  {journal} {\bibinfo  {journal} {Phys. Rev. A}\ }\textbf {\bibinfo {volume} {84}},\ \bibinfo {pages} {034301} (\bibinfo {year} {2011})}\BibitemShut {NoStop}%
\bibitem [{\citenamefont {Vallone}\ \emph {et~al.}(2014)\citenamefont {Vallone}, \citenamefont {Marangon}, \citenamefont {Tomasin},\ and\ \citenamefont {Villoresi}}]{Vallone2014}%
  \BibitemOpen
  \bibfield  {author} {\bibinfo {author} {\bibfnamefont {G.}~\bibnamefont {Vallone}}, \bibinfo {author} {\bibfnamefont {D.~G.}\ \bibnamefont {Marangon}}, \bibinfo {author} {\bibfnamefont {M.}~\bibnamefont {Tomasin}},\ and\ \bibinfo {author} {\bibfnamefont {P.}~\bibnamefont {Villoresi}},\ }\bibfield  {title} {\bibinfo {title} {{Quantum randomness certified by the uncertainty principle}},\ }\href {https://doi.org/10.1103/PhysRevA.90.052327} {\bibfield  {journal} {\bibinfo  {journal} {Phys. Rev. A}\ }\textbf {\bibinfo {volume} {90}},\ \bibinfo {pages} {052327} (\bibinfo {year} {2014})}\BibitemShut {NoStop}%
\bibitem [{\citenamefont {Lunghi}\ \emph {et~al.}(2015)\citenamefont {Lunghi}, \citenamefont {Brask}, \citenamefont {Lim}, \citenamefont {Lavigne}, \citenamefont {Bowles}, \citenamefont {Martin}, \citenamefont {Zbinden},\ and\ \citenamefont {Brunner}}]{Lunghi2015}%
  \BibitemOpen
  \bibfield  {author} {\bibinfo {author} {\bibfnamefont {T.}~\bibnamefont {Lunghi}}, \bibinfo {author} {\bibfnamefont {J.~B.}\ \bibnamefont {Brask}}, \bibinfo {author} {\bibfnamefont {C.~C.~W.}\ \bibnamefont {Lim}}, \bibinfo {author} {\bibfnamefont {Q.}~\bibnamefont {Lavigne}}, \bibinfo {author} {\bibfnamefont {J.}~\bibnamefont {Bowles}}, \bibinfo {author} {\bibfnamefont {A.}~\bibnamefont {Martin}}, \bibinfo {author} {\bibfnamefont {H.}~\bibnamefont {Zbinden}},\ and\ \bibinfo {author} {\bibfnamefont {N.}~\bibnamefont {Brunner}},\ }\bibfield  {title} {\bibinfo {title} {Self-testing quantum random number generator},\ }\href {https://doi.org/10.1103/PhysRevLett.114.150501} {\bibfield  {journal} {\bibinfo  {journal} {Phys. Rev. Lett.}\ }\textbf {\bibinfo {volume} {114}},\ \bibinfo {pages} {150501} (\bibinfo {year} {2015})}\BibitemShut {NoStop}%
\bibitem [{\citenamefont {Chaturvedi}\ and\ \citenamefont {Banik}(2015)}]{Chaturvedi2015}%
  \BibitemOpen
  \bibfield  {author} {\bibinfo {author} {\bibfnamefont {A.}~\bibnamefont {Chaturvedi}}\ and\ \bibinfo {author} {\bibfnamefont {M.}~\bibnamefont {Banik}},\ }\bibfield  {title} {\bibinfo {title} {Measurement-device-independent randomness from local entangled states},\ }\href {https://doi.org/10.1209/0295-5075/112/30003} {\bibfield  {journal} {\bibinfo  {journal} {{EPL} (Europhysics Letters)}\ }\textbf {\bibinfo {volume} {112}},\ \bibinfo {pages} {30003} (\bibinfo {year} {2015})}\BibitemShut {NoStop}%
\bibitem [{\citenamefont {Cao}\ \emph {et~al.}(2015)\citenamefont {Cao}, \citenamefont {Zhou},\ and\ \citenamefont {Ma}}]{Cao2015}%
  \BibitemOpen
  \bibfield  {author} {\bibinfo {author} {\bibfnamefont {Z.}~\bibnamefont {Cao}}, \bibinfo {author} {\bibfnamefont {H.}~\bibnamefont {Zhou}},\ and\ \bibinfo {author} {\bibfnamefont {X.}~\bibnamefont {Ma}},\ }\bibfield  {title} {\bibinfo {title} {{Loss-tolerant measurement-device-independent quantum random number generation}},\ }\href {https://doi.org/10.1088/1367-2630/17/12/125011} {\bibfield  {journal} {\bibinfo  {journal} {New J. Phys.}\ }\textbf {\bibinfo {volume} {17}},\ \bibinfo {pages} {125011} (\bibinfo {year} {2015})}\BibitemShut {NoStop}%
\bibitem [{\citenamefont {Nie}\ \emph {et~al.}(2016)\citenamefont {Nie}, \citenamefont {Guan}, \citenamefont {Zhou}, \citenamefont {Zhang}, \citenamefont {Ma}, \citenamefont {Zhang},\ and\ \citenamefont {Pan}}]{Nie2016}%
  \BibitemOpen
  \bibfield  {author} {\bibinfo {author} {\bibfnamefont {Y.-Q.}\ \bibnamefont {Nie}}, \bibinfo {author} {\bibfnamefont {J.-Y.}\ \bibnamefont {Guan}}, \bibinfo {author} {\bibfnamefont {H.}~\bibnamefont {Zhou}}, \bibinfo {author} {\bibfnamefont {Q.}~\bibnamefont {Zhang}}, \bibinfo {author} {\bibfnamefont {X.}~\bibnamefont {Ma}}, \bibinfo {author} {\bibfnamefont {J.}~\bibnamefont {Zhang}},\ and\ \bibinfo {author} {\bibfnamefont {J.-W.}\ \bibnamefont {Pan}},\ }\bibfield  {title} {\bibinfo {title} {Experimental measurement-device-independent quantum random-number generation},\ }\href {https://doi.org/10.1103/PhysRevA.94.060301} {\bibfield  {journal} {\bibinfo  {journal} {Phys. Rev. A}\ }\textbf {\bibinfo {volume} {94}},\ \bibinfo {pages} {060301} (\bibinfo {year} {2016})}\BibitemShut {NoStop}%
\bibitem [{\citenamefont {Cao}\ \emph {et~al.}(2016)\citenamefont {Cao}, \citenamefont {Zhou}, \citenamefont {Yuan},\ and\ \citenamefont {Ma}}]{Cao2016}%
  \BibitemOpen
  \bibfield  {author} {\bibinfo {author} {\bibfnamefont {Z.}~\bibnamefont {Cao}}, \bibinfo {author} {\bibfnamefont {H.}~\bibnamefont {Zhou}}, \bibinfo {author} {\bibfnamefont {X.}~\bibnamefont {Yuan}},\ and\ \bibinfo {author} {\bibfnamefont {X.}~\bibnamefont {Ma}},\ }\bibfield  {title} {\bibinfo {title} {{Source-Independent Quantum Random Number Generation}},\ }\href {https://doi.org/10.1103/PhysRevX.6.011020} {\bibfield  {journal} {\bibinfo  {journal} {Phys. Rev. X}\ }\textbf {\bibinfo {volume} {6}},\ \bibinfo {pages} {011020} (\bibinfo {year} {2016})}\BibitemShut {NoStop}%
\bibitem [{\citenamefont {Xu}\ \emph {et~al.}(2016)\citenamefont {Xu}, \citenamefont {Shapiro},\ and\ \citenamefont {Wong}}]{Xu2016}%
  \BibitemOpen
  \bibfield  {author} {\bibinfo {author} {\bibfnamefont {F.}~\bibnamefont {Xu}}, \bibinfo {author} {\bibfnamefont {J.~H.}\ \bibnamefont {Shapiro}},\ and\ \bibinfo {author} {\bibfnamefont {F.~N.~C.}\ \bibnamefont {Wong}},\ }\bibfield  {title} {\bibinfo {title} {Experimental fast quantum random number generation using high-dimensional entanglement with entropy monitoring},\ }\href {https://doi.org/10.1364/OPTICA.3.001266} {\bibfield  {journal} {\bibinfo  {journal} {Optica}\ }\textbf {\bibinfo {volume} {3}},\ \bibinfo {pages} {1266} (\bibinfo {year} {2016})}\BibitemShut {NoStop}%
\bibitem [{\citenamefont {Marangon}\ \emph {et~al.}(2017)\citenamefont {Marangon}, \citenamefont {Vallone},\ and\ \citenamefont {Villoresi}}]{Marangon2017}%
  \BibitemOpen
  \bibfield  {author} {\bibinfo {author} {\bibfnamefont {D.~G.}\ \bibnamefont {Marangon}}, \bibinfo {author} {\bibfnamefont {G.}~\bibnamefont {Vallone}},\ and\ \bibinfo {author} {\bibfnamefont {P.}~\bibnamefont {Villoresi}},\ }\bibfield  {title} {\bibinfo {title} {Source-device-independent ultrafast quantum random number generation},\ }\href {https://doi.org/10.1103/PhysRevLett.118.060503} {\bibfield  {journal} {\bibinfo  {journal} {Phys. Rev. Lett.}\ }\textbf {\bibinfo {volume} {118}},\ \bibinfo {pages} {060503} (\bibinfo {year} {2017})}\BibitemShut {NoStop}%
\bibitem [{\citenamefont {Brask}\ \emph {et~al.}(2017)\citenamefont {Brask}, \citenamefont {Martin}, \citenamefont {Esposito}, \citenamefont {Houlmann}, \citenamefont {Bowles}, \citenamefont {Zbinden},\ and\ \citenamefont {Brunner}}]{Brask2017mhz}%
  \BibitemOpen
  \bibfield  {author} {\bibinfo {author} {\bibfnamefont {J.~B.}\ \bibnamefont {Brask}}, \bibinfo {author} {\bibfnamefont {A.}~\bibnamefont {Martin}}, \bibinfo {author} {\bibfnamefont {W.}~\bibnamefont {Esposito}}, \bibinfo {author} {\bibfnamefont {R.}~\bibnamefont {Houlmann}}, \bibinfo {author} {\bibfnamefont {J.}~\bibnamefont {Bowles}}, \bibinfo {author} {\bibfnamefont {H.}~\bibnamefont {Zbinden}},\ and\ \bibinfo {author} {\bibfnamefont {N.}~\bibnamefont {Brunner}},\ }\bibfield  {title} {\bibinfo {title} {{Megahertz-Rate Semi-Device-Independent Quantum Random Number Generators Based on Unambiguous State Discrimination}},\ }\href {https://doi.org/10.1103/PhysRevApplied.7.054018} {\bibfield  {journal} {\bibinfo  {journal} {Phys. Rev. Appl.}\ }\textbf {\bibinfo {volume} {7}},\ \bibinfo {pages} {054018} (\bibinfo {year} {2017})}\BibitemShut {NoStop}%
\bibitem [{\citenamefont {Avesani}\ \emph {et~al.}(2018)\citenamefont {Avesani}, \citenamefont {Marangon}, \citenamefont {Vallone},\ and\ \citenamefont {Villoresi}}]{Avesani2018}%
  \BibitemOpen
  \bibfield  {author} {\bibinfo {author} {\bibfnamefont {M.}~\bibnamefont {Avesani}}, \bibinfo {author} {\bibfnamefont {D.~G.}\ \bibnamefont {Marangon}}, \bibinfo {author} {\bibfnamefont {G.}~\bibnamefont {Vallone}},\ and\ \bibinfo {author} {\bibfnamefont {P.}~\bibnamefont {Villoresi}},\ }\bibfield  {title} {\bibinfo {title} {Source-device-independent heterodyne-based quantum random number generator at 17 gbps},\ }\href {https://doi.org/10.1038/s41467-018-07585-0} {\bibfield  {journal} {\bibinfo  {journal} {Nature Communications}\ }\textbf {\bibinfo {volume} {9}},\ \bibinfo {pages} {5365} (\bibinfo {year} {2018})}\BibitemShut {NoStop}%
\bibitem [{\citenamefont {Michel}\ \emph {et~al.}(2019)\citenamefont {Michel}, \citenamefont {Haw}, \citenamefont {Marangon}, \citenamefont {Thearle}, \citenamefont {Vallone}, \citenamefont {Villoresi}, \citenamefont {Lam},\ and\ \citenamefont {Assad}}]{Michel2019}%
  \BibitemOpen
  \bibfield  {author} {\bibinfo {author} {\bibfnamefont {T.}~\bibnamefont {Michel}}, \bibinfo {author} {\bibfnamefont {J.~Y.}\ \bibnamefont {Haw}}, \bibinfo {author} {\bibfnamefont {D.~G.}\ \bibnamefont {Marangon}}, \bibinfo {author} {\bibfnamefont {O.}~\bibnamefont {Thearle}}, \bibinfo {author} {\bibfnamefont {G.}~\bibnamefont {Vallone}}, \bibinfo {author} {\bibfnamefont {P.}~\bibnamefont {Villoresi}}, \bibinfo {author} {\bibfnamefont {P.~K.}\ \bibnamefont {Lam}},\ and\ \bibinfo {author} {\bibfnamefont {S.~M.}\ \bibnamefont {Assad}},\ }\bibfield  {title} {\bibinfo {title} {Real-time source-independent quantum random-number generator with squeezed states},\ }\href {https://doi.org/10.1103/PhysRevApplied.12.034017} {\bibfield  {journal} {\bibinfo  {journal} {Phys. Rev. Applied}\ }\textbf {\bibinfo {volume} {12}},\ \bibinfo {pages} {034017} (\bibinfo {year} {2019})}\BibitemShut {NoStop}%
\bibitem [{\citenamefont {Rusca}\ \emph {et~al.}(2019)\citenamefont {Rusca}, \citenamefont {van Himbeeck}, \citenamefont {Martin}, \citenamefont {Brask}, \citenamefont {Shi}, \citenamefont {Pironio}, \citenamefont {Brunner},\ and\ \citenamefont {Zbinden}}]{Rusca2019}%
  \BibitemOpen
  \bibfield  {author} {\bibinfo {author} {\bibfnamefont {D.}~\bibnamefont {Rusca}}, \bibinfo {author} {\bibfnamefont {T.}~\bibnamefont {van Himbeeck}}, \bibinfo {author} {\bibfnamefont {A.}~\bibnamefont {Martin}}, \bibinfo {author} {\bibfnamefont {J.~B.}\ \bibnamefont {Brask}}, \bibinfo {author} {\bibfnamefont {W.}~\bibnamefont {Shi}}, \bibinfo {author} {\bibfnamefont {S.}~\bibnamefont {Pironio}}, \bibinfo {author} {\bibfnamefont {N.}~\bibnamefont {Brunner}},\ and\ \bibinfo {author} {\bibfnamefont {H.}~\bibnamefont {Zbinden}},\ }\bibfield  {title} {\bibinfo {title} {Self-testing quantum random-number generator based on an energy bound},\ }\href {https://doi.org/10.1103/PhysRevA.100.062338} {\bibfield  {journal} {\bibinfo  {journal} {Phys. Rev. A}\ }\textbf {\bibinfo {volume} {100}},\ \bibinfo {pages} {062338} (\bibinfo {year} {2019})}\BibitemShut {NoStop}%
\bibitem [{\citenamefont {Drahi}\ \emph {et~al.}(2020)\citenamefont {Drahi}, \citenamefont {Walk}, \citenamefont {Hoban}, \citenamefont {Fedorov}, \citenamefont {Shakhovoy}, \citenamefont {Feimov}, \citenamefont {Kurochkin}, \citenamefont {Kolthammer}, \citenamefont {Nunn}, \citenamefont {Barrett},\ and\ \citenamefont {Walmsley}}]{Drahi2020}%
  \BibitemOpen
  \bibfield  {author} {\bibinfo {author} {\bibfnamefont {D.}~\bibnamefont {Drahi}}, \bibinfo {author} {\bibfnamefont {N.}~\bibnamefont {Walk}}, \bibinfo {author} {\bibfnamefont {M.~J.}\ \bibnamefont {Hoban}}, \bibinfo {author} {\bibfnamefont {A.~K.}\ \bibnamefont {Fedorov}}, \bibinfo {author} {\bibfnamefont {R.}~\bibnamefont {Shakhovoy}}, \bibinfo {author} {\bibfnamefont {A.}~\bibnamefont {Feimov}}, \bibinfo {author} {\bibfnamefont {Y.}~\bibnamefont {Kurochkin}}, \bibinfo {author} {\bibfnamefont {W.~S.}\ \bibnamefont {Kolthammer}}, \bibinfo {author} {\bibfnamefont {J.}~\bibnamefont {Nunn}}, \bibinfo {author} {\bibfnamefont {J.}~\bibnamefont {Barrett}},\ and\ \bibinfo {author} {\bibfnamefont {I.~A.}\ \bibnamefont {Walmsley}},\ }\bibfield  {title} {\bibinfo {title} {Certified quantum random numbers from untrusted light},\ }\href {https://doi.org/10.1103/PhysRevX.10.041048} {\bibfield  {journal} {\bibinfo  {journal} {Phys. Rev. X}\ }\textbf {\bibinfo {volume} {10}},\ \bibinfo {pages} {041048} (\bibinfo {year}
  {2020})}\BibitemShut {NoStop}%
\bibitem [{\citenamefont {Rusca}\ \emph {et~al.}(2020)\citenamefont {Rusca}, \citenamefont {Tebyanian}, \citenamefont {Martin},\ and\ \citenamefont {Zbinden}}]{Rusca2020}%
  \BibitemOpen
  \bibfield  {author} {\bibinfo {author} {\bibfnamefont {D.}~\bibnamefont {Rusca}}, \bibinfo {author} {\bibfnamefont {H.}~\bibnamefont {Tebyanian}}, \bibinfo {author} {\bibfnamefont {A.}~\bibnamefont {Martin}},\ and\ \bibinfo {author} {\bibfnamefont {H.}~\bibnamefont {Zbinden}},\ }\bibfield  {title} {\bibinfo {title} {Fast self-testing quantum random number generator based on homodyne detection},\ }\href {https://doi.org/10.1063/5.0011479} {\bibfield  {journal} {\bibinfo  {journal} {Applied Physics Letters}\ }\textbf {\bibinfo {volume} {116}},\ \bibinfo {pages} {264004} (\bibinfo {year} {2020})}\BibitemShut {NoStop}%
\bibitem [{\citenamefont {Mironowicz}\ \emph {et~al.}(2021)\citenamefont {Mironowicz}, \citenamefont {Cañas}, \citenamefont {Cariñe}, \citenamefont {Gómez}, \citenamefont {Barra}, \citenamefont {Cabello}, \citenamefont {Xavier}, \citenamefont {Lima},\ and\ \citenamefont {Pawłowski}}]{Mironowicz2021}%
  \BibitemOpen
  \bibfield  {author} {\bibinfo {author} {\bibfnamefont {P.}~\bibnamefont {Mironowicz}}, \bibinfo {author} {\bibfnamefont {G.}~\bibnamefont {Cañas}}, \bibinfo {author} {\bibfnamefont {J.}~\bibnamefont {Cariñe}}, \bibinfo {author} {\bibfnamefont {E.~S.}\ \bibnamefont {Gómez}}, \bibinfo {author} {\bibfnamefont {J.~F.}\ \bibnamefont {Barra}}, \bibinfo {author} {\bibfnamefont {A.}~\bibnamefont {Cabello}}, \bibinfo {author} {\bibfnamefont {G.~B.}\ \bibnamefont {Xavier}}, \bibinfo {author} {\bibfnamefont {G.}~\bibnamefont {Lima}},\ and\ \bibinfo {author} {\bibfnamefont {M.}~\bibnamefont {Pawłowski}},\ }\bibfield  {title} {\bibinfo {title} {Quantum randomness protected against detection loophole attacks},\ }\href {https://doi.org/10.1007/s11128-020-02948-3} {\bibfield  {journal} {\bibinfo  {journal} {Quantum Information Processing}\ }\textbf {\bibinfo {volume} {20}},\ \bibinfo {pages} {39} (\bibinfo {year} {2021})}\BibitemShut {NoStop}%
\bibitem [{\citenamefont {Wang}\ \emph {et~al.}(2023)\citenamefont {Wang}, \citenamefont {Primaatmaja}, \citenamefont {Ng}, \citenamefont {Haw}, \citenamefont {Ho}, \citenamefont {Zhang}, \citenamefont {Zhang},\ and\ \citenamefont {Lim}}]{Wang2023}%
  \BibitemOpen
  \bibfield  {author} {\bibinfo {author} {\bibfnamefont {C.}~\bibnamefont {Wang}}, \bibinfo {author} {\bibfnamefont {I.~W.}\ \bibnamefont {Primaatmaja}}, \bibinfo {author} {\bibfnamefont {H.~J.}\ \bibnamefont {Ng}}, \bibinfo {author} {\bibfnamefont {J.~Y.}\ \bibnamefont {Haw}}, \bibinfo {author} {\bibfnamefont {R.}~\bibnamefont {Ho}}, \bibinfo {author} {\bibfnamefont {J.}~\bibnamefont {Zhang}}, \bibinfo {author} {\bibfnamefont {G.}~\bibnamefont {Zhang}},\ and\ \bibinfo {author} {\bibfnamefont {C.}~\bibnamefont {Lim}},\ }\bibfield  {title} {\bibinfo {title} {Provably-secure quantum randomness expansion with uncharacterised homodyne detection},\ }\href {https://doi.org/10.1038/s41467-022-35556-z} {\bibfield  {journal} {\bibinfo  {journal} {Nature Communications}\ }\textbf {\bibinfo {volume} {14}},\ \bibinfo {pages} {316} (\bibinfo {year} {2023})}\BibitemShut {NoStop}%
\bibitem [{\citenamefont {Gabriel}\ \emph {et~al.}(2010)\citenamefont {Gabriel}, \citenamefont {Wittmann}, \citenamefont {Sych}, \citenamefont {Dong}, \citenamefont {Mauerer}, \citenamefont {Andersen}, \citenamefont {Marquardt},\ and\ \citenamefont {Leuchs}}]{Gabriel2010}%
  \BibitemOpen
  \bibfield  {author} {\bibinfo {author} {\bibfnamefont {C.}~\bibnamefont {Gabriel}}, \bibinfo {author} {\bibfnamefont {C.}~\bibnamefont {Wittmann}}, \bibinfo {author} {\bibfnamefont {D.}~\bibnamefont {Sych}}, \bibinfo {author} {\bibfnamefont {R.}~\bibnamefont {Dong}}, \bibinfo {author} {\bibfnamefont {W.}~\bibnamefont {Mauerer}}, \bibinfo {author} {\bibfnamefont {U.~L.}\ \bibnamefont {Andersen}}, \bibinfo {author} {\bibfnamefont {C.}~\bibnamefont {Marquardt}},\ and\ \bibinfo {author} {\bibfnamefont {G.}~\bibnamefont {Leuchs}},\ }\bibfield  {title} {\bibinfo {title} {A generator for unique quantum random numbers based on vacuum states},\ }\href {https://doi.org/10.1038/nphoton.2010.197} {\bibfield  {journal} {\bibinfo  {journal} {Nature Photonics}\ }\textbf {\bibinfo {volume} {4}},\ \bibinfo {pages} {711} (\bibinfo {year} {2010})}\BibitemShut {NoStop}%
\bibitem [{\citenamefont {Symul}\ \emph {et~al.}(2011)\citenamefont {Symul}, \citenamefont {Assad},\ and\ \citenamefont {Lam}}]{Symul2011}%
  \BibitemOpen
  \bibfield  {author} {\bibinfo {author} {\bibfnamefont {T.}~\bibnamefont {Symul}}, \bibinfo {author} {\bibfnamefont {S.~M.}\ \bibnamefont {Assad}},\ and\ \bibinfo {author} {\bibfnamefont {P.~K.}\ \bibnamefont {Lam}},\ }\bibfield  {title} {\bibinfo {title} {Real time demonstration of high bitrate quantum random number generation with coherent laser light},\ }\href {https://doi.org/10.1063/1.3597793} {\bibfield  {journal} {\bibinfo  {journal} {Applied Physics Letters}\ }\textbf {\bibinfo {volume} {98}},\ \bibinfo {pages} {231103} (\bibinfo {year} {2011})}\BibitemShut {NoStop}%
\bibitem [{\citenamefont {Zhang}\ \emph {et~al.}(2016)\citenamefont {Zhang}, \citenamefont {Nie}, \citenamefont {Zhou}, \citenamefont {Liang}, \citenamefont {Ma}, \citenamefont {Zhang},\ and\ \citenamefont {Pan}}]{Zhang2016}%
  \BibitemOpen
  \bibfield  {author} {\bibinfo {author} {\bibfnamefont {X.-G.}\ \bibnamefont {Zhang}}, \bibinfo {author} {\bibfnamefont {Y.-Q.}\ \bibnamefont {Nie}}, \bibinfo {author} {\bibfnamefont {H.}~\bibnamefont {Zhou}}, \bibinfo {author} {\bibfnamefont {H.}~\bibnamefont {Liang}}, \bibinfo {author} {\bibfnamefont {X.}~\bibnamefont {Ma}}, \bibinfo {author} {\bibfnamefont {J.}~\bibnamefont {Zhang}},\ and\ \bibinfo {author} {\bibfnamefont {J.-W.}\ \bibnamefont {Pan}},\ }\bibfield  {title} {\bibinfo {title} {Note: Fully integrated 3.2 gbps quantum random number generator with real-time extraction},\ }\href {https://doi.org/10.1063/1.4958663} {\bibfield  {journal} {\bibinfo  {journal} {Review of Scientific Instruments}\ }\textbf {\bibinfo {volume} {87}},\ \bibinfo {pages} {076102} (\bibinfo {year} {2016})}\BibitemShut {NoStop}%
\bibitem [{\citenamefont {Huang}\ and\ \citenamefont {Zhou}(2019)}]{Huang2019}%
  \BibitemOpen
  \bibfield  {author} {\bibinfo {author} {\bibfnamefont {L.}~\bibnamefont {Huang}}\ and\ \bibinfo {author} {\bibfnamefont {H.}~\bibnamefont {Zhou}},\ }\bibfield  {title} {\bibinfo {title} {Integrated gbps quantum random number generator with real-time extraction based on homodyne detection},\ }\href {https://doi.org/10.1364/JOSAB.36.00B130} {\bibfield  {journal} {\bibinfo  {journal} {J. Opt. Soc. Am. B}\ }\textbf {\bibinfo {volume} {36}},\ \bibinfo {pages} {B130} (\bibinfo {year} {2019})}\BibitemShut {NoStop}%
\bibitem [{\citenamefont {Gehring}\ \emph {et~al.}(2021)\citenamefont {Gehring}, \citenamefont {Lupo}, \citenamefont {Kordts}, \citenamefont {Solar~Nikolic}, \citenamefont {Jain}, \citenamefont {Rydberg}, \citenamefont {Pedersen}, \citenamefont {Pirandola},\ and\ \citenamefont {Andersen}}]{Gehring2021}%
  \BibitemOpen
  \bibfield  {author} {\bibinfo {author} {\bibfnamefont {T.}~\bibnamefont {Gehring}}, \bibinfo {author} {\bibfnamefont {C.}~\bibnamefont {Lupo}}, \bibinfo {author} {\bibfnamefont {A.}~\bibnamefont {Kordts}}, \bibinfo {author} {\bibfnamefont {D.}~\bibnamefont {Solar~Nikolic}}, \bibinfo {author} {\bibfnamefont {N.}~\bibnamefont {Jain}}, \bibinfo {author} {\bibfnamefont {T.}~\bibnamefont {Rydberg}}, \bibinfo {author} {\bibfnamefont {T.~B.}\ \bibnamefont {Pedersen}}, \bibinfo {author} {\bibfnamefont {S.}~\bibnamefont {Pirandola}},\ and\ \bibinfo {author} {\bibfnamefont {U.~L.}\ \bibnamefont {Andersen}},\ }\bibfield  {title} {\bibinfo {title} {Homodyne-based quantum random number generator at 2.9 gbps secure against quantum side-information},\ }\href {https://doi.org/10.1038/s41467-020-20813-w} {\bibfield  {journal} {\bibinfo  {journal} {Nature Communications}\ }\textbf {\bibinfo {volume} {12}},\ \bibinfo {pages} {605} (\bibinfo {year} {2021})}\BibitemShut {NoStop}%
\bibitem [{\citenamefont {Tomamichel}\ \emph {et~al.}(2011)\citenamefont {Tomamichel}, \citenamefont {Schaffner}, \citenamefont {Smith},\ and\ \citenamefont {Renner}}]{tomamichel2011}%
  \BibitemOpen
  \bibfield  {author} {\bibinfo {author} {\bibfnamefont {M.}~\bibnamefont {Tomamichel}}, \bibinfo {author} {\bibfnamefont {C.}~\bibnamefont {Schaffner}}, \bibinfo {author} {\bibfnamefont {A.}~\bibnamefont {Smith}},\ and\ \bibinfo {author} {\bibfnamefont {R.}~\bibnamefont {Renner}},\ }\bibfield  {title} {\bibinfo {title} {Leftover hashing against quantum side information},\ }\href {https://doi.org/10.1109/TIT.2011.2158473} {\bibfield  {journal} {\bibinfo  {journal} {IEEE Transactions on Information Theory}\ }\textbf {\bibinfo {volume} {57}},\ \bibinfo {pages} {5524} (\bibinfo {year} {2011})}\BibitemShut {NoStop}%
\bibitem [{\citenamefont {Carceller}\ \emph {et~al.}(2025)\citenamefont {Carceller}, \citenamefont {Faria}, \citenamefont {Liu}, \citenamefont {Sguerso}, \citenamefont {Andersen}, \citenamefont {Neergaard-Nielsen},\ and\ \citenamefont {Brask}}]{carceller2025_improving}%
  \BibitemOpen
  \bibfield  {author} {\bibinfo {author} {\bibfnamefont {C.~R.~i.}\ \bibnamefont {Carceller}}, \bibinfo {author} {\bibfnamefont {L.~N.}\ \bibnamefont {Faria}}, \bibinfo {author} {\bibfnamefont {Z.-H.}\ \bibnamefont {Liu}}, \bibinfo {author} {\bibfnamefont {N.}~\bibnamefont {Sguerso}}, \bibinfo {author} {\bibfnamefont {U.~L.}\ \bibnamefont {Andersen}}, \bibinfo {author} {\bibfnamefont {J.~S.}\ \bibnamefont {Neergaard-Nielsen}},\ and\ \bibinfo {author} {\bibfnamefont {J.~B.}\ \bibnamefont {Brask}},\ }\bibfield  {title} {\bibinfo {title} {Improving semi-device-independent randomness certification by entropy accumulation},\ }\href {https://doi.org/10.1103/dwdv-89bj} {\bibfield  {journal} {\bibinfo  {journal} {Phys. Rev. A}\ }\textbf {\bibinfo {volume} {112}},\ \bibinfo {pages} {022430} (\bibinfo {year} {2025})}\BibitemShut {NoStop}%
\bibitem [{\citenamefont {Arnon-Friedman}\ \emph {et~al.}(2018)\citenamefont {Arnon-Friedman}, \citenamefont {Dupuis}, \citenamefont {Fawzi}, \citenamefont {Renner},\ and\ \citenamefont {Vidick}}]{ArnonFriedman2018}%
  \BibitemOpen
  \bibfield  {author} {\bibinfo {author} {\bibfnamefont {R.}~\bibnamefont {Arnon-Friedman}}, \bibinfo {author} {\bibfnamefont {F.}~\bibnamefont {Dupuis}}, \bibinfo {author} {\bibfnamefont {O.}~\bibnamefont {Fawzi}}, \bibinfo {author} {\bibfnamefont {R.}~\bibnamefont {Renner}},\ and\ \bibinfo {author} {\bibfnamefont {T.}~\bibnamefont {Vidick}},\ }\bibfield  {title} {\bibinfo {title} {Practical device-independent quantum cryptography via entropy accumulation},\ }\href {https://doi.org/10.1038/s41467-017-02307-4} {\bibfield  {journal} {\bibinfo  {journal} {Nature Communications}\ }\textbf {\bibinfo {volume} {9}},\ \bibinfo {pages} {459} (\bibinfo {year} {2018})}\BibitemShut {NoStop}%
\bibitem [{\citenamefont {Dupuis}\ \emph {et~al.}(2020)\citenamefont {Dupuis}, \citenamefont {Fawzi},\ and\ \citenamefont {Renner}}]{Dupuis2020}%
  \BibitemOpen
  \bibfield  {author} {\bibinfo {author} {\bibfnamefont {F.}~\bibnamefont {Dupuis}}, \bibinfo {author} {\bibfnamefont {O.}~\bibnamefont {Fawzi}},\ and\ \bibinfo {author} {\bibfnamefont {R.}~\bibnamefont {Renner}},\ }\bibfield  {title} {\bibinfo {title} {Entropy accumulation},\ }\href {https://doi.org/10.1007/s00220-020-03839-5} {\bibfield  {journal} {\bibinfo  {journal} {Communications in Mathematical Physics}\ }\textbf {\bibinfo {volume} {379}},\ \bibinfo {pages} {867} (\bibinfo {year} {2020})}\BibitemShut {NoStop}%
\bibitem [{\citenamefont {Metger}\ \emph {et~al.}(2022)\citenamefont {Metger}, \citenamefont {Fawzi}, \citenamefont {Sutter},\ and\ \citenamefont {Renner}}]{metger2022}%
  \BibitemOpen
  \bibfield  {author} {\bibinfo {author} {\bibfnamefont {T.}~\bibnamefont {Metger}}, \bibinfo {author} {\bibfnamefont {O.}~\bibnamefont {Fawzi}}, \bibinfo {author} {\bibfnamefont {D.}~\bibnamefont {Sutter}},\ and\ \bibinfo {author} {\bibfnamefont {R.}~\bibnamefont {Renner}},\ }\bibfield  {title} {\bibinfo {title} {Generalised entropy accumulation},\ }in\ \href {https://doi.org/10.1109/FOCS54457.2022.00085} {\emph {\bibinfo {booktitle} {2022 IEEE 63rd Annual Symposium on Foundations of Computer Science (FOCS)}}}\ (\bibinfo {year} {2022})\ pp.\ \bibinfo {pages} {844--850}\BibitemShut {NoStop}%
\bibitem [{\citenamefont {Metger}\ and\ \citenamefont {Renner}(2023)}]{Metger2023}%
  \BibitemOpen
  \bibfield  {author} {\bibinfo {author} {\bibfnamefont {T.}~\bibnamefont {Metger}}\ and\ \bibinfo {author} {\bibfnamefont {R.}~\bibnamefont {Renner}},\ }\bibfield  {title} {\bibinfo {title} {Security for quantum key distribution from generalised entropy accumulation},\ }\href@noop {} {\bibfield  {journal} {\bibinfo  {journal} {Nat. Commun.}\ }\textbf {\bibinfo {volume} {14}},\ \bibinfo {pages} {5272} (\bibinfo {year} {2023})}\BibitemShut {NoStop}%
\bibitem [{\citenamefont {Brown}\ \emph {et~al.}(2024)\citenamefont {Brown}, \citenamefont {Fawzi},\ and\ \citenamefont {Fawzi}}]{Brown2024}%
  \BibitemOpen
  \bibfield  {author} {\bibinfo {author} {\bibfnamefont {P.}~\bibnamefont {Brown}}, \bibinfo {author} {\bibfnamefont {H.}~\bibnamefont {Fawzi}},\ and\ \bibinfo {author} {\bibfnamefont {O.}~\bibnamefont {Fawzi}},\ }\bibfield  {title} {\bibinfo {title} {Device-independent lower bounds on the conditional von {N}eumann entropy},\ }\href {https://doi.org/10.22331/q-2024-08-27-1445} {\bibfield  {journal} {\bibinfo  {journal} {{Quantum}}\ }\textbf {\bibinfo {volume} {8}},\ \bibinfo {pages} {1445} (\bibinfo {year} {2024})}\BibitemShut {NoStop}%
\bibitem [{\citenamefont {Roch~i Carceller}\ \emph {et~al.}(2025)\citenamefont {Roch~i Carceller}, \citenamefont {Pauwels}, \citenamefont {Pironio},\ and\ \citenamefont {Tavakoli}}]{carceller2025_photon}%
  \BibitemOpen
  \bibfield  {author} {\bibinfo {author} {\bibfnamefont {C.}~\bibnamefont {Roch~i Carceller}}, \bibinfo {author} {\bibfnamefont {J.}~\bibnamefont {Pauwels}}, \bibinfo {author} {\bibfnamefont {S.}~\bibnamefont {Pironio}},\ and\ \bibinfo {author} {\bibfnamefont {A.}~\bibnamefont {Tavakoli}},\ }\bibfield  {title} {\bibinfo {title} {Prepare-and-measure scenarios with photon-number constraints},\ }\href {https://doi.org/10.1103/glty-kkbp} {\bibfield  {journal} {\bibinfo  {journal} {Phys. Rev. Lett.}\ }\textbf {\bibinfo {volume} {135}},\ \bibinfo {pages} {140802} (\bibinfo {year} {2025})}\BibitemShut {NoStop}%
\bibitem [{\citenamefont {Koßmann}\ and\ \citenamefont {Schwonnek}(2026)}]{kossmann2026}%
  \BibitemOpen
  \bibfield  {author} {\bibinfo {author} {\bibfnamefont {G.}~\bibnamefont {Koßmann}}\ and\ \bibinfo {author} {\bibfnamefont {R.}~\bibnamefont {Schwonnek}},\ }\bibfield  {title} {\bibinfo {title} {Optimising the relative entropy under semidefinite constraints},\ }\href {https://doi.org/10.1038/s41534-026-01184-4} {\bibfield  {journal} {\bibinfo  {journal} {npj Quantum Information}\ }\textbf {\bibinfo {volume} {12}},\ \bibinfo {pages} {23} (\bibinfo {year} {2026})}\BibitemShut {NoStop}%
\bibitem [{Note1()}]{Note1}%
  \BibitemOpen
  \bibinfo {note} {We note that, in the absence of noise and for a tomographically complete set of probe states (or equivalently a fully characterised POVM), the certifiable randomness will be $\Delta $ bits. However, the coherent-state probes considered here are not tomographically complete and even for a complete set, convergence towards maximal randomness with SNR is slow}\BibitemShut {NoStop}%
\bibitem [{\citenamefont {Avesani}\ \emph {et~al.}(2021)\citenamefont {Avesani}, \citenamefont {Tebyanian}, \citenamefont {Villoresi},\ and\ \citenamefont {Vallone}}]{Avesani2021}%
  \BibitemOpen
  \bibfield  {author} {\bibinfo {author} {\bibfnamefont {M.}~\bibnamefont {Avesani}}, \bibinfo {author} {\bibfnamefont {H.}~\bibnamefont {Tebyanian}}, \bibinfo {author} {\bibfnamefont {P.}~\bibnamefont {Villoresi}},\ and\ \bibinfo {author} {\bibfnamefont {G.}~\bibnamefont {Vallone}},\ }\bibfield  {title} {\bibinfo {title} {Semi-device-independent heterodyne-based quantum random-number generator},\ }\href {https://doi.org/10.1103/PhysRevApplied.15.034034} {\bibfield  {journal} {\bibinfo  {journal} {Phys. Rev. Appl.}\ }\textbf {\bibinfo {volume} {15}},\ \bibinfo {pages} {034034} (\bibinfo {year} {2021})}\BibitemShut {NoStop}%
\bibitem [{\citenamefont {Nie}\ \emph {et~al.}(2024)\citenamefont {Nie}, \citenamefont {Zhou}, \citenamefont {Bai}, \citenamefont {Xu}, \citenamefont {Ma}, \citenamefont {Zhang},\ and\ \citenamefont {Pan}}]{Nie2024}%
  \BibitemOpen
  \bibfield  {author} {\bibinfo {author} {\bibfnamefont {Y.-Q.}\ \bibnamefont {Nie}}, \bibinfo {author} {\bibfnamefont {H.}~\bibnamefont {Zhou}}, \bibinfo {author} {\bibfnamefont {B.}~\bibnamefont {Bai}}, \bibinfo {author} {\bibfnamefont {Q.}~\bibnamefont {Xu}}, \bibinfo {author} {\bibfnamefont {X.}~\bibnamefont {Ma}}, \bibinfo {author} {\bibfnamefont {J.}~\bibnamefont {Zhang}},\ and\ \bibinfo {author} {\bibfnamefont {J.-W.}\ \bibnamefont {Pan}},\ }\bibfield  {title} {\bibinfo {title} {Measurement-device-independent quantum random number generation over 23 {Mbps} with imperfect single-photon sources},\ }\href {https://doi.org/10.1088/2058-9565/ad34f4} {\bibfield  {journal} {\bibinfo  {journal} {Quantum Science and Technology}\ }\textbf {\bibinfo {volume} {9}},\ \bibinfo {pages} {025024} (\bibinfo {year} {2024})}\BibitemShut {NoStop}%
\bibitem [{\citenamefont {Argillander}\ \emph {et~al.}(2025)\citenamefont {Argillander}, \citenamefont {Spegel-Lexne}, \citenamefont {Clason}, \citenamefont {Dieguez}, \citenamefont {Pawłowski}, \citenamefont {Chaturvedi},\ and\ \citenamefont {Xavier}}]{Argillander2025}%
  \BibitemOpen
  \bibfield  {author} {\bibinfo {author} {\bibfnamefont {J.}~\bibnamefont {Argillander}}, \bibinfo {author} {\bibfnamefont {D.}~\bibnamefont {Spegel-Lexne}}, \bibinfo {author} {\bibfnamefont {M.}~\bibnamefont {Clason}}, \bibinfo {author} {\bibfnamefont {P.~R.}\ \bibnamefont {Dieguez}}, \bibinfo {author} {\bibfnamefont {M.}~\bibnamefont {Pawłowski}}, \bibinfo {author} {\bibfnamefont {A.}~\bibnamefont {Chaturvedi}},\ and\ \bibinfo {author} {\bibfnamefont {G.~B.}\ \bibnamefont {Xavier}},\ }\bibfield  {title} {\bibinfo {title} {High-dimensional detection-loophole-free measurement-device-independent quantum random number generator},\ }\bibfield  {journal} {\bibinfo  {journal} {arXiv:2510.06317}\ }\href {https://doi.org/10.48550/arXiv.2510.06317} {10.48550/arXiv.2510.06317} (\bibinfo {year} {2025})\BibitemShut {NoStop}%
\bibitem [{\citenamefont {Sabatini}\ \emph {et~al.}(2026)\citenamefont {Sabatini}, \citenamefont {i~Carceller}, \citenamefont {Bertapelle}, \citenamefont {Peri}, \citenamefont {Gualandi}, \citenamefont {Bolaños}, \citenamefont {Piétri}, \citenamefont {Corrielli}, \citenamefont {Villoresi}, \citenamefont {Vallone}, \citenamefont {Osellame}, \citenamefont {Tavakoli},\ and\ \citenamefont {Avesani}}]{Sabatini2026}%
  \BibitemOpen
  \bibfield  {author} {\bibinfo {author} {\bibfnamefont {M.}~\bibnamefont {Sabatini}}, \bibinfo {author} {\bibfnamefont {C.~R.}\ \bibnamefont {i~Carceller}}, \bibinfo {author} {\bibfnamefont {T.}~\bibnamefont {Bertapelle}}, \bibinfo {author} {\bibfnamefont {A.}~\bibnamefont {Peri}}, \bibinfo {author} {\bibfnamefont {G.}~\bibnamefont {Gualandi}}, \bibinfo {author} {\bibfnamefont {M.~R.}\ \bibnamefont {Bolaños}}, \bibinfo {author} {\bibfnamefont {Y.}~\bibnamefont {Piétri}}, \bibinfo {author} {\bibfnamefont {G.}~\bibnamefont {Corrielli}}, \bibinfo {author} {\bibfnamefont {P.}~\bibnamefont {Villoresi}}, \bibinfo {author} {\bibfnamefont {G.}~\bibnamefont {Vallone}}, \bibinfo {author} {\bibfnamefont {R.}~\bibnamefont {Osellame}}, \bibinfo {author} {\bibfnamefont {A.}~\bibnamefont {Tavakoli}},\ and\ \bibinfo {author} {\bibfnamefont {M.}~\bibnamefont {Avesani}},\ }\bibfield  {title} {\bibinfo {title} {Efficient energy-constrained semi-device-independent {QRNG} with an integrated heterodyne receiver},\ }\bibfield
  {journal} {\bibinfo  {journal} {arXiv:2606.23231}\ }\href {https://doi.org/10.48550/arXiv.2606.23231} {10.48550/arXiv.2606.23231} (\bibinfo {year} {2026})\BibitemShut {NoStop}%
\bibitem [{\citenamefont {Bian}\ \emph {et~al.}(2025)\citenamefont {Bian}, \citenamefont {Yang}, \citenamefont {Jiang}, \citenamefont {Huang}, \citenamefont {Su}, \citenamefont {Yu}, \citenamefont {Zhang}, \citenamefont {Zhang},\ and\ \citenamefont {Xu}}]{Bian2025}%
  \BibitemOpen
  \bibfield  {author} {\bibinfo {author} {\bibfnamefont {Y.}~\bibnamefont {Bian}}, \bibinfo {author} {\bibfnamefont {J.}~\bibnamefont {Yang}}, \bibinfo {author} {\bibfnamefont {H.}~\bibnamefont {Jiang}}, \bibinfo {author} {\bibfnamefont {W.}~\bibnamefont {Huang}}, \bibinfo {author} {\bibfnamefont {Q.}~\bibnamefont {Su}}, \bibinfo {author} {\bibfnamefont {S.}~\bibnamefont {Yu}}, \bibinfo {author} {\bibfnamefont {L.}~\bibnamefont {Zhang}}, \bibinfo {author} {\bibfnamefont {Y.}~\bibnamefont {Zhang}},\ and\ \bibinfo {author} {\bibfnamefont {B.}~\bibnamefont {Xu}},\ }\bibfield  {title} {\bibinfo {title} {20 {Gbps} real-time source-independent quantum random number generator based on a silicon photonic chip},\ }\href {https://doi.org/10.1364/OL.544982} {\bibfield  {journal} {\bibinfo  {journal} {Opt. Lett.}\ }\textbf {\bibinfo {volume} {50}},\ \bibinfo {pages} {1216} (\bibinfo {year} {2025})}\BibitemShut {NoStop}%
\bibitem [{\citenamefont {Bertapelle}\ \emph {et~al.}(2025)\citenamefont {Bertapelle}, \citenamefont {Avesani}, \citenamefont {Santamato}, \citenamefont {Montanaro}, \citenamefont {Chiesa}, \citenamefont {Rotta}, \citenamefont {Artiglia}, \citenamefont {Sorianello}, \citenamefont {Testa}, \citenamefont {Angelis}, \citenamefont {Contestabile}, \citenamefont {Vallone}, \citenamefont {Romagnoli},\ and\ \citenamefont {Villoresi}}]{Bertapelle2025}%
  \BibitemOpen
  \bibfield  {author} {\bibinfo {author} {\bibfnamefont {T.}~\bibnamefont {Bertapelle}}, \bibinfo {author} {\bibfnamefont {M.}~\bibnamefont {Avesani}}, \bibinfo {author} {\bibfnamefont {A.}~\bibnamefont {Santamato}}, \bibinfo {author} {\bibfnamefont {A.}~\bibnamefont {Montanaro}}, \bibinfo {author} {\bibfnamefont {M.}~\bibnamefont {Chiesa}}, \bibinfo {author} {\bibfnamefont {D.}~\bibnamefont {Rotta}}, \bibinfo {author} {\bibfnamefont {M.}~\bibnamefont {Artiglia}}, \bibinfo {author} {\bibfnamefont {V.}~\bibnamefont {Sorianello}}, \bibinfo {author} {\bibfnamefont {F.}~\bibnamefont {Testa}}, \bibinfo {author} {\bibfnamefont {G.~D.}\ \bibnamefont {Angelis}}, \bibinfo {author} {\bibfnamefont {G.}~\bibnamefont {Contestabile}}, \bibinfo {author} {\bibfnamefont {G.}~\bibnamefont {Vallone}}, \bibinfo {author} {\bibfnamefont {M.}~\bibnamefont {Romagnoli}},\ and\ \bibinfo {author} {\bibfnamefont {P.}~\bibnamefont {Villoresi}},\ }\bibfield  {title} {\bibinfo {title} {High-speed source-device-independent quantum random
  number generator on a chip},\ }\href {https://doi.org/10.1364/OPTICAQ.529746} {\bibfield  {journal} {\bibinfo  {journal} {Optica Quantum}\ }\textbf {\bibinfo {volume} {3}},\ \bibinfo {pages} {111} (\bibinfo {year} {2025})}\BibitemShut {NoStop}%
\bibitem [{\citenamefont {Cizauskas}\ \emph {et~al.}(2025)\citenamefont {Cizauskas}, \citenamefont {Tebyanian}, \citenamefont {Fox}, \citenamefont {Bayer}, \citenamefont {Assmann},\ and\ \citenamefont {Greilich}}]{Cizauskas2025}%
  \BibitemOpen
  \bibfield  {author} {\bibinfo {author} {\bibfnamefont {M.}~\bibnamefont {Cizauskas}}, \bibinfo {author} {\bibfnamefont {H.}~\bibnamefont {Tebyanian}}, \bibinfo {author} {\bibfnamefont {M.}~\bibnamefont {Fox}}, \bibinfo {author} {\bibfnamefont {M.}~\bibnamefont {Bayer}}, \bibinfo {author} {\bibfnamefont {M.}~\bibnamefont {Assmann}},\ and\ \bibinfo {author} {\bibfnamefont {A.}~\bibnamefont {Greilich}},\ }\bibfield  {title} {\bibinfo {title} {{33 Gbit/s source-device-independent quantum random number generator based on heterodyne detection with real-time FPGA-integrated extraction}},\ }\bibfield  {journal} {\bibinfo  {journal} {arXiv:2512.07319}\ }\href {https://doi.org/https://doi.org/10.48550/arXiv.2512.07319} {https://doi.org/10.48550/arXiv.2512.07319} (\bibinfo {year} {2025})\BibitemShut {NoStop}%
\bibitem [{\citenamefont {Kincaid}\ \emph {et~al.}(2025)\citenamefont {Kincaid}, \citenamefont {Marinis}, \citenamefont {Testa}, \citenamefont {Andriolli},\ and\ \citenamefont {Contestabile}}]{Kincaid2025}%
  \BibitemOpen
  \bibfield  {author} {\bibinfo {author} {\bibfnamefont {P.~S.}\ \bibnamefont {Kincaid}}, \bibinfo {author} {\bibfnamefont {L.~D.}\ \bibnamefont {Marinis}}, \bibinfo {author} {\bibfnamefont {F.}~\bibnamefont {Testa}}, \bibinfo {author} {\bibfnamefont {N.}~\bibnamefont {Andriolli}},\ and\ \bibinfo {author} {\bibfnamefont {G.}~\bibnamefont {Contestabile}},\ }\bibfield  {title} {\bibinfo {title} {{Source-device-independent monolithically integrated QRNG in a black box with generation rate in excess of 30 Gbit/s}},\ }\bibfield  {journal} {\bibinfo  {journal} {arXiv:2510.18700}\ }\href {https://doi.org/https://doi.org/10.48550/arXiv.2510.18700} {https://doi.org/10.48550/arXiv.2510.18700} (\bibinfo {year} {2025})\BibitemShut {NoStop}%
\bibitem [{\citenamefont {Golub}(1973)}]{golub1973}%
  \BibitemOpen
  \bibfield  {author} {\bibinfo {author} {\bibfnamefont {G.~H.}\ \bibnamefont {Golub}},\ }\bibfield  {title} {\bibinfo {title} {Some modified matrix eigenvalue problems},\ }\href {https://doi.org/10.1137/1015032} {\bibfield  {journal} {\bibinfo  {journal} {SIAM Review}\ }\textbf {\bibinfo {volume} {15}},\ \bibinfo {pages} {318} (\bibinfo {year} {1973})},\ \Eprint {https://arxiv.org/abs/https://doi.org/10.1137/1015032} {https://doi.org/10.1137/1015032} \BibitemShut {NoStop}%
\bibitem [{\citenamefont {Bancal}\ \emph {et~al.}(2014)\citenamefont {Bancal}, \citenamefont {Sheridan},\ and\ \citenamefont {Scarani}}]{bancal2014more}%
  \BibitemOpen
  \bibfield  {author} {\bibinfo {author} {\bibfnamefont {J.-D.}\ \bibnamefont {Bancal}}, \bibinfo {author} {\bibfnamefont {L.}~\bibnamefont {Sheridan}},\ and\ \bibinfo {author} {\bibfnamefont {V.}~\bibnamefont {Scarani}},\ }\bibfield  {title} {\bibinfo {title} {More randomness from the same data},\ }\href@noop {} {\bibfield  {journal} {\bibinfo  {journal} {New Journal of Physics}\ }\textbf {\bibinfo {volume} {16}},\ \bibinfo {pages} {033011} (\bibinfo {year} {2014})}\BibitemShut {NoStop}%
\bibitem [{\citenamefont {Renner}\ and\ \citenamefont {Wolf}(2004)}]{Renner2004}%
  \BibitemOpen
  \bibfield  {author} {\bibinfo {author} {\bibfnamefont {R.}~\bibnamefont {Renner}}\ and\ \bibinfo {author} {\bibfnamefont {S.}~\bibnamefont {Wolf}},\ }\bibfield  {title} {\bibinfo {title} {{Smooth R{\'{e}}nyi entropy and applications}},\ }in\ \href@noop {} {\emph {\bibinfo {booktitle} {Proc. Int. Symp. Inf. Theory}}}\ (\bibinfo {year} {2004})\ p.\ \bibinfo {pages} {233}\BibitemShut {NoStop}%
\bibitem [{\citenamefont {Diamond}\ and\ \citenamefont {Boyd}(2016)}]{diamond2016cvxpy}%
  \BibitemOpen
  \bibfield  {author} {\bibinfo {author} {\bibfnamefont {S.}~\bibnamefont {Diamond}}\ and\ \bibinfo {author} {\bibfnamefont {S.}~\bibnamefont {Boyd}},\ }\bibfield  {title} {\bibinfo {title} {{CVXPY}: {A} {P}ython-embedded modeling language for convex optimization},\ }\href {http://jmlr.org/papers/v17/15-408.html} {\bibfield  {journal} {\bibinfo  {journal} {Journal of Machine Learning Research}\ }\textbf {\bibinfo {volume} {17}},\ \bibinfo {pages} {1} (\bibinfo {year} {2016})}\BibitemShut {NoStop}%
\bibitem [{\citenamefont {Agrawal}\ \emph {et~al.}(2018)\citenamefont {Agrawal}, \citenamefont {Verschueren}, \citenamefont {Diamond},\ and\ \citenamefont {Boyd}}]{agrawal2018rewriting}%
  \BibitemOpen
  \bibfield  {author} {\bibinfo {author} {\bibfnamefont {A.}~\bibnamefont {Agrawal}}, \bibinfo {author} {\bibfnamefont {R.}~\bibnamefont {Verschueren}}, \bibinfo {author} {\bibfnamefont {S.}~\bibnamefont {Diamond}},\ and\ \bibinfo {author} {\bibfnamefont {S.}~\bibnamefont {Boyd}},\ }\bibfield  {title} {\bibinfo {title} {A rewriting system for convex optimization problems},\ }\href@noop {} {\bibfield  {journal} {\bibinfo  {journal} {Journal of Control and Decision}\ }\textbf {\bibinfo {volume} {5}},\ \bibinfo {pages} {42} (\bibinfo {year} {2018})}\BibitemShut {NoStop}%
\bibitem [{\citenamefont {APS}(2024)}]{Mosek2024}%
  \BibitemOpen
  \bibfield  {author} {\bibinfo {author} {\bibfnamefont {M.}~\bibnamefont {APS}},\ }\href@noop {} {\emph {\bibinfo {title} {MOSEK Optimizer API for Python 10.2.8}}} (\bibinfo {year} {2024})\BibitemShut {NoStop}%
\bibitem [{\citenamefont {Virtanen}\ \emph {et~al.}(2020)\citenamefont {Virtanen}, \citenamefont {Gommers}, \citenamefont {Oliphant}, \citenamefont {Haberland}, \citenamefont {Reddy}, \citenamefont {Cournapeau}, \citenamefont {Burovski}, \citenamefont {Peterson}, \citenamefont {Weckesser}, \citenamefont {Bright}, \citenamefont {{van der Walt}}, \citenamefont {Brett}, \citenamefont {Wilson}, \citenamefont {Millman}, \citenamefont {Mayorov}, \citenamefont {Nelson}, \citenamefont {Jones}, \citenamefont {Kern}, \citenamefont {Larson}, \citenamefont {Carey}, \citenamefont {Polat}, \citenamefont {Feng}, \citenamefont {Moore}, \citenamefont {{VanderPlas}}, \citenamefont {Laxalde}, \citenamefont {Perktold}, \citenamefont {Cimrman}, \citenamefont {Henriksen}, \citenamefont {Quintero}, \citenamefont {Harris}, \citenamefont {Archibald}, \citenamefont {Ribeiro}, \citenamefont {Pedregosa}, \citenamefont {{van Mulbregt}},\ and\ \citenamefont {{SciPy 1.0 Contributors}}}]{2020SciPy-NMeth}%
  \BibitemOpen
  \bibfield  {author} {\bibinfo {author} {\bibfnamefont {P.}~\bibnamefont {Virtanen}}, \bibinfo {author} {\bibfnamefont {R.}~\bibnamefont {Gommers}}, \bibinfo {author} {\bibfnamefont {T.~E.}\ \bibnamefont {Oliphant}}, \bibinfo {author} {\bibfnamefont {M.}~\bibnamefont {Haberland}}, \bibinfo {author} {\bibfnamefont {T.}~\bibnamefont {Reddy}}, \bibinfo {author} {\bibfnamefont {D.}~\bibnamefont {Cournapeau}}, \bibinfo {author} {\bibfnamefont {E.}~\bibnamefont {Burovski}}, \bibinfo {author} {\bibfnamefont {P.}~\bibnamefont {Peterson}}, \bibinfo {author} {\bibfnamefont {W.}~\bibnamefont {Weckesser}}, \bibinfo {author} {\bibfnamefont {J.}~\bibnamefont {Bright}}, \bibinfo {author} {\bibfnamefont {S.~J.}\ \bibnamefont {{van der Walt}}}, \bibinfo {author} {\bibfnamefont {M.}~\bibnamefont {Brett}}, \bibinfo {author} {\bibfnamefont {J.}~\bibnamefont {Wilson}}, \bibinfo {author} {\bibfnamefont {K.~J.}\ \bibnamefont {Millman}}, \bibinfo {author} {\bibfnamefont {N.}~\bibnamefont {Mayorov}}, \bibinfo {author} {\bibfnamefont
  {A.~R.~J.}\ \bibnamefont {Nelson}}, \bibinfo {author} {\bibfnamefont {E.}~\bibnamefont {Jones}}, \bibinfo {author} {\bibfnamefont {R.}~\bibnamefont {Kern}}, \bibinfo {author} {\bibfnamefont {E.}~\bibnamefont {Larson}}, \bibinfo {author} {\bibfnamefont {C.~J.}\ \bibnamefont {Carey}}, \bibinfo {author} {\bibfnamefont {{\.I}.}~\bibnamefont {Polat}}, \bibinfo {author} {\bibfnamefont {Y.}~\bibnamefont {Feng}}, \bibinfo {author} {\bibfnamefont {E.~W.}\ \bibnamefont {Moore}}, \bibinfo {author} {\bibfnamefont {J.}~\bibnamefont {{VanderPlas}}}, \bibinfo {author} {\bibfnamefont {D.}~\bibnamefont {Laxalde}}, \bibinfo {author} {\bibfnamefont {J.}~\bibnamefont {Perktold}}, \bibinfo {author} {\bibfnamefont {R.}~\bibnamefont {Cimrman}}, \bibinfo {author} {\bibfnamefont {I.}~\bibnamefont {Henriksen}}, \bibinfo {author} {\bibfnamefont {E.~A.}\ \bibnamefont {Quintero}}, \bibinfo {author} {\bibfnamefont {C.~R.}\ \bibnamefont {Harris}}, \bibinfo {author} {\bibfnamefont {A.~M.}\ \bibnamefont {Archibald}}, \bibinfo {author}
  {\bibfnamefont {A.~H.}\ \bibnamefont {Ribeiro}}, \bibinfo {author} {\bibfnamefont {F.}~\bibnamefont {Pedregosa}}, \bibinfo {author} {\bibfnamefont {P.}~\bibnamefont {{van Mulbregt}}},\ and\ \bibinfo {author} {\bibnamefont {{SciPy 1.0 Contributors}}},\ }\bibfield  {title} {\bibinfo {title} {{{SciPy} 1.0: Fundamental Algorithms for Scientific Computing in Python}},\ }\href {https://doi.org/10.1038/s41592-019-0686-2} {\bibfield  {journal} {\bibinfo  {journal} {Nature Methods}\ }\textbf {\bibinfo {volume} {17}},\ \bibinfo {pages} {261} (\bibinfo {year} {2020})}\BibitemShut {NoStop}%
\end{thebibliography}%

\onecolumngrid
\appendix
\counterwithin{figure}{section} 

\section{SDP relaxation for lower-bounding the von Neumann entropy} \label{app.shannon}

In this section we derive from scratch the SDP relaxation presented in the main text used to bound the conditional von Neumann entropy of the measurement output. We begin presenting the correlations in our scenario when an eavesdropper (Eve) with the goal of guessing the measurement outcome is added to the picture. We continue presenting the conditional von Neumann entropy, and how it reduces to the conditional Shannon entropy of the measurement output. We finally derive an SDP to bound the conditional Shannon entropy during the generation rounds, and its dual in the form of the SDP relaxation presented in the main text.

\subsection{Scenario and correlations}

Consider the tri-partite scenario in which three particles are prepared into a quantum state $\Psi_{ABE}$, and are shared among three parties, namely Alice, Bob and Eve. In Alice's side, she will privately select a classical input $i$ and perform a transformation described by the completely positive and trace preserving (CPTP) map $\Omega_i^{A}$. Afterwards, she sends her particle to Bob, who will jointly measure it with his own particle from the original shared state using the measurement $\{M_k^{AB}\}$ and receive $k$ as his measurement outcome. Finally, Eve will perform the measurement $\{E_\lambda^{E}\}$ to her particle in her own laboratory and receive $\lambda$ as measurement result. This process is repeated multiple times, producing the observable correlations in the form of the following input-output probabilities,
\begin{align} \label{eq:pbe_probs}
p(k,\lambda|i) = \Tr\left[\left(\Omega_i^{A}\otimes \mathbb{I}_{BE}\right)\left[\Psi_{ABE}\right]\left(M_k^{AB}\otimes E_\lambda^{E}\right)\right] \ .
\end{align}
We now consider a measurement-device-independent framework, in which all state preparations coming out from Alice's device are trusted and known to be the pure states $\rho_i = \ketbra{\psi_i}{\psi_i}$. In this framework, the tri-partite state after Alice's operation becomes $\left(\Omega_i^{A}\otimes \mathbb{I}_{BE}\right)\left[\Psi_{ABE}\right] = \rho_i^{A} \otimes \sigma_{BE}$. Inserting these in \eqref{eq:pbe_probs} amounts to
\begin{align}\label{eq:joint_bl}
p(k,\lambda|\rho_i) &= \Tr\left[\left(\rho_i^{A}\otimes \sigma_{BE}\right)\left(M_k^{AB}\otimes E_\lambda^{E}\right)\right] = q_{\lambda}\Tr\left[\left(\rho_{i}^{A}\otimes \sigma_{B}^\lambda\right)M_{k}^{AB}\right] \ , 
\end{align}
where we defined
\begin{align}
q_{\lambda}=\Tr\left[\sigma_{BE}\left(\mathbb{I}_B\otimes E_{\lambda}^{E}\right)\right] \quad \text{and} \quad \sigma_{B}^\lambda = \frac{\Tr_E\left[\sigma_{BE}\left(\mathbb{I}_B\otimes E_{\lambda}^{E}\right)\right]}{q_{\lambda}}
\end{align} 
to be the steered assemblage in Bob's device by Eve's measurement.

The resulting correlations are interpreted as follows. During every round, Eve chooses a hidden variable $\lambda$ with probability $q_{\lambda}$ and remotely prepares a state $\sigma_B^\lambda$ in Bob's device. Globally, the whole initial quantum state can be written as
\begin{align}\label{eq:psi_ABE_int}
\Psi_{ABE} =  \rho^{A}\otimes \sum_\lambda q_{\lambda} \sigma_{B}^\lambda \otimes \ketbra{\lambda}{\lambda} \ .
\end{align} 
Then, Alice performs her local transformation $\Omega_i^{A}$ and produces the state $\rho_i^A$ which she sends to Bob. The global state now becomes,
\begin{align}
\left(\Omega_i^{A}\otimes \mathbb{I}_{BE}\right)\left[\Psi_{ABE}\right] =  \rho^{A}_i\otimes \sum_\lambda q_{\lambda} \sigma_{B}^\lambda \otimes \ketbra{\lambda}{\lambda} \ .
\end{align} 
Then, Bob applies a measurement $M_k^{AB}$ to both the state $\rho_i^{A}$ that Alice produced and the state $\sigma_B^\lambda$ that Eve produced. Eve is only left with her hidden variable $\lambda$ in a classical register. The observable correlations for Alice and Bob, i.e.~averaged over the hidden variable $\lambda$, then become
\begin{align}
p(k|\rho_i) = \sum_\lambda q_{\lambda} \Tr\left[\left(\rho_{i}^{A}\otimes \sigma_{B}^\lambda\right)M_{k}^{AB}\right] \ ,
\end{align}
and the joint distribution is simply $p(k,\lambda|\rho_i) = q_{\lambda}p_\lambda(k|\rho_i)$, where $p_\lambda(k|\rho_i) = \Tr\left[\left(\rho_{i}^{A}\otimes \sigma_{B}^\lambda\right)M_{k}^{AB}\right]$.

\subsection{Relating the von Neumann and Shannon entropies}

\subsubsection{Definition in our framework: reduction to the Shannon entropy}

The von Neumann entropy of a system $B$ conditioned on Eve's side information $E$ during generation rounds (in our case, when $\rho_0$ is prepared) can be expressed through the unnormalized relative entropy $D(\rho||\sigma)=\Tr\left[\rho\left(\log_2\rho-\log_2\sigma\right)\right]$ as,
\begin{align} \label{eq:app_VN_rel}
H(B|E) = -D\left(\tilde{\Psi}_{0}||\mathbb{I}\otimes \bar{\sigma}_{0}\right) \ ,
\end{align}
where $\tilde{\Psi}_{0}$ is the classical-quantum state resulting after Bob's measurement and $\bar{\sigma}_{0}$ is the averaged state left to Eve after Bob's measurement. Concretely, these can be written as
\begin{align}
\tilde{\Psi}_{i} = \sum_k \ketbra{k}{k}\otimes \sigma_{k|i}
\end{align}
with
\begin{align}
\sigma_{k|i}&=\Tr_{AB}\left[\left(\Omega_{i}^{A}\otimes \mathbb{I}_{BE}\right)\left[\Psi_{ABE}\right]\left(M_{k}^{AB}\otimes\mathbb{I}_{E}\right)\right] = \sum_{\lambda}q_{\lambda}\Tr\left[\left(\rho_i^A\otimes \sigma_B^\lambda\right)M_{k}^{AB}\right]\ketbra{\lambda}{\lambda} \ , 
\end{align}
where we inserted the state $\Psi_{ABE}$ from \eqref{eq:psi_ABE_int} and $\bar{\sigma}_{i}=\sum_k \sigma_{k|i}$. Substituting in \eqref{eq:app_VN_rel}, one ends up with
\begin{align} \label{eq:vn_entropy}
H(B|E) &= - \Tr\left[ \sum_k \ketbra{k}{k}\otimes\sigma_{k|0}\left(\log_2\left(\sum_k \ketbra{k}{k}\otimes\sigma_{k|0}\right) - \log_2\left(\sum_k \ketbra{k}{k}\otimes \sum_{k'} \sigma_{k'|0}\right)\right) \right] \\
&= - \sum_k \Tr\left[ \sigma_{k|0}\left(\log_2\sigma_{k|0} - \log_2 \sum_{k'} \sigma_{k'|0}\right) \right] = -\sum_\lambda q_{\lambda}\sum_k\Tr\left[\left(\rho_{0}^A\otimes \sigma_B^\lambda\right)M_{k}^{AB}\right] \log_2 \Tr\left[\left(\rho_{0}^A\otimes \sigma_B^\lambda\right)M_{k}^{AB}\right]   \nonumber \\
&= -\sum_\lambda q_{\lambda} \sum_k p_\lambda(k|\rho_0) \log_2 p_\lambda(k|\rho_0)  \ , \nonumber
\end{align}
for $p_\lambda(k|\rho_i)=\Tr\left[\left(\rho_i^A\otimes \sigma_B^\lambda\right)M_{k}^{AB}\right]=\Tr\left[\rho_i\hat{\Pi}_k^\lambda\right]$, where we defined $\hat{\Pi}_k^\lambda = \Tr_{B}\left[\left(\mathbb{I}_A\otimes \sigma_{B}^\lambda\right)M_k^{AB}\right]$. Thus, at the end of the day, the von Neumann entropy reduces to the conditional Shannon entropy.

\subsubsection{Bounding the Shannon entropy}

To bound the Shannon entropy, we use the method from Ref.~\cite{carceller2025_photon}, which consists in an alternative solution parallel to BFF \cite{Brown2024}. First, we express the base 2 logarithm $\log_2(u) = \log(u)/\log(2)$ in \eqref{eq:vn_entropy} in terms of the natural logarithm $\log(u)$ and use the integral representation
\begin{equation}
    \log(u) = \int_{0}^{1} \frac{u-1}{t(u-1)+1}\, \dd t , \quad \text{with} \quad -\log(u)=\log\left(\frac{1}{u}\right),
\end{equation}
to write the Shannon entropy as 
\begin{equation}
    H(B|E) = \sum_\lambda q_{\lambda}\sum_{k} p_\lambda(k|\rho_0) \frac{1}{\log 2} \int_{0}^{1} \frac{1-p_\lambda(k|\rho_0)}{t(1-p_\lambda(k|\rho_0)) + p_\lambda(k|\rho_0)} \, \dd t\,.
\end{equation}
Then, we discretize the integral using the Gauss-Radau quadratures to obtain the bound
\begin{equation}
H(B|E) \geq \sum_\lambda q_{\lambda}\sum_{k} p_\lambda(k|\rho_0)  \sum_{j=1}^{m-1} \frac{w_j}{\log 2} \frac{1-p_\lambda(k|\rho_0)}{t_j(1-p_\lambda(k|\rho_0)) + p_\lambda(k|\rho_0)} ,
\end{equation}
where $\omega_j$ and $t_j$ are the weights and nodes of the Gauss-Radau quadrature, which can be straightforwardly obtained with a simple algorithm as detailed in Ref.~\cite{golub1973}. After some algebra, one obtains
\begin{equation}
H(B|E) \geq \sum_\lambda q_{\lambda}\sum_{k} p_\lambda(k|\rho_0) \sum_{j=1}^{m-1} \frac{w_j}{t_j \log 2} \left( 1 - \frac{p_\lambda(k|\rho_0)}{t_j(1-p_\lambda(k|\rho_0)) + p_\lambda(k|\rho_0)} \right),
\end{equation}
which can be rewritten as
\begin{equation}\label{eq:shannon_bound_fbi}
H(B|E) \geq c_m +  \sum_\lambda q_{\lambda}\sum_{j, k} \tau_j \left(- \frac{p_\lambda(k|\rho_0)^2}{t_j(1-p_\lambda(k|\rho_0)) + p_\lambda(k|\rho_0)} \right) = c_m + \sum_{j, k} \tau_j f_{k,j},
\end{equation}
for $\tau_j := \frac{\omega_j}{t_j \log 2}$ and $c_m:=\sum_{j=1}^{m-1}\tau_j$, and where we defined
\begin{equation}
f_{k,j}:= -\sum_\lambda q_{\lambda}\frac{p_\lambda(k|\rho_0)^2}{t_j(1-p_\lambda(k|\rho_0)) + p_\lambda(k|\rho_0)}.
\end{equation}
The above expressions $f_{k,j}$ are a convex combination of rationals $f = -\frac{p^2}{t(1-p)+p}$, which often arise as the minimum of a quadratic function $f = \min_z g(z) = \min_z \left(az^2 + bz\right)$. The minimum of $g(z)$ occurs at $z^* = - \frac{b}{2a}$ if $a>0$, with $g(z^*) = -\frac{b^2}{4a}$. Thus, taking $a = p(1-t)+t$ and $b=2p$, we write $f_{k,j}$ as
\begin{equation}
f_{k,j} = \sum_\lambda q_{\lambda} \min_{z_{k,j}^{\lambda}} p_\lambda(k|\rho_0) \left[2z_{k,j}^{\lambda} + (1-t_j) (z_{k,j}^{\lambda})^2 \right] + t_j (z_{k,j}^{\lambda})^2  \ .
\end{equation}
Inserting this expression in the bound \eqref{eq:shannon_bound_fbi}, and using $p_\lambda(k|\rho_0) = \Tr\left[\rho_{0} \hat{\Pi}_{k}^\lambda\right]$, we find 
\begin{align} \label{eq:shann_inf}
	H(B|E) \geq c_m + \sum_{j=1}^{m-1}\sum_{k} \tau_j \sum_\lambda q_\lambda  \min_{z_{k,j}^{\lambda}} \left\{ \Tr\left[\rho_{0} \hat{\Pi}_{k}^{\lambda}\right] \left[2z_{k,j}^{\lambda} + (1-t_j) (z_{k,j}^{\lambda})^2 \right] + t_j (z_{k,j}^{\lambda})^2 \right\}
\end{align}
which is the same bound on the Shannon entropy one would obtain through a direct application of BFF \cite{Brown2024}.

\subsection{A single SDP relaxation to bound the Shannon entropy}

\subsubsection{Primal SDP relaxation}

We now compute the lower bound on the right-hand-side of \eqref{eq:shann_inf} using a semidefinite program relaxation. In order to do so, our aim will be to group up the variables $z_{k,j}^{\lambda}$ with the uncharacterised measurement operators $\hat{\Pi}_k^{\lambda}$ in order to linearise the problem. To do so, we will sample a list of operators $O=\{\hat{\Pi}_k^{\lambda}, z_{j,a}^{\lambda} \hat{\Pi}_k^{\lambda} , (z_{j,a}^{\lambda})^2 \hat{\Pi}_k^{\lambda}, \ldots \}$ up to a certain order on $z_{j,a}^{\lambda}$ (the sub-index $k$ was replaced by the dummy $a$ to denote the independence from the POVM element sub-index $\hat{\Pi}_k^{\lambda}$). With these operators we build the following block-matrix
\begin{align}
	G^{ja}_{k} = \sum_\lambda q_\lambda\begin{pmatrix}
	\hat{\Pi}_k^{\lambda} & z_{a,j}^{\lambda} \hat{\Pi}_k^{\lambda} & (z_{a,j}^{\lambda})^2 \hat{\Pi}_k^{\lambda} & \cdots \\
	z_{a,j}^{\lambda} \hat{\Pi}_k^{\lambda} & (z_{a,j}^{\lambda})^2 \hat{\Pi}_k^{\lambda} & (z_{a,j}^{\lambda})^3 \hat{\Pi}_k & \cdots \\
	\vdots & \vdots & \vdots & \ddots
	\end{pmatrix} \succeq 0 \ ,
\end{align}
with monomials up to certain order $r$ in $z_{a,j}^{\lambda}$, which is positive-semidefinite by construction. Moreover, note that each block $\left(G^{ja}_{k}\right)_{u,v} = \sum_\lambda q_\lambda (z_{j,a}^{\lambda})^{u+v} \hat{\Pi}_k^{\lambda}$ for $u+v>0$ satisfies
\begin{align}
	&\sum_k \left(G^{ja}_{k}\right)_{u,v} = \frac{1}{D}\Tr\left[ \sum_k \left(G^{ja}_{k}\right)_{u,v} \right] \mathbb{I} \ ,
\end{align}
where $D$ is the dimension of the POVM elements $\hat{\Pi}_k^{\lambda}$. We can now re-write the Shannon entropy bound in \eqref{eq:shann_inf} and minimise over the defined operators above, first defining
\begin{align}
	&f_{j,a} = \sum_k \sum_{u,v} c_{u,v}^{j}(a,k) \Tr\left[\rho_{0} \left(G^{ja}_{k}\right)_{u,v} \right]
\end{align}
for 
\begin{align}
	c_{u,v}^{j}(a,k) = 2\delta_{k,a}\delta_{u+v,1}+(1-t_j)\delta_{k,a}\delta_{u+v,2} + t_j \delta_{u+v,2} \ .
\end{align}
Note we can render the minimisation of $f_{j,a}$ from \eqref{eq:shann_inf} independently for each $j$. Therefore, we can omit the indexing $j$ on each variable, as it will be reset after each iteration. This leads to the following primal semidefinite program,
\begin{equation}\label{eq:shannon_primal} 
\begin{aligned}
    H(B|E) \geq c_m + \sum_{j=1}^{m-1} \ \underset{G^{a}_{k}}{\text{min}} & \quad \tau_j \sum_{a,k} \sum_{u,v} c_{u,v}^{j}(a,k) \Tr\left[\rho_{0} \left(G^{a}_{k}\right)_{u,v} \right], \\
    \text{subject to} & \quad G^{a}_{k} \succeq 0, \\
    & \quad \Tr\left[ \rho_i \left(G^{a}_{k}\right)_{0,0} \right] = p(k|\rho_i) ,  \\
    & \quad \sum_{k}\left(G^{a}_{k}\right)_{0,0} = \mathbb{I} , \\
    & \quad \sum_k \left(G^{a}_{k}\right)_{u,v} = \frac{1}{D}\Tr\left[ \sum_k \left(G^{a}_{k}\right)_{u,v} \right] \mathbb{I} .
\end{aligned}
\end{equation}
Any set of $rD\times rD$ matrices $G^{a}_{k}$ satisfying the primal constraints above represents a valid lower bound on the Shannon entropy. Here $r$ is the maximum order of the monomials $(z_{a,j}^{\lambda})^r \hat{\Pi}_k^{\lambda}$ taken to build $G^{a}_{k}$, and $2r$ defines the level of the SDP relaxation.

\subsubsection{Dual SDP relaxation}

We now proceed to derive the dual form of the SDP relaxation above. The Lagrangian corresponding to the complete optimisation reads $\mathcal{L} = c_{m} + \sum_j \mathcal{L}_j$, for
\begin{equation}
\begin{aligned}
\mathcal{L}_j = & \sum_{a,k} \red{\sum_{u,v}}\tau_j \left( 2\delta_{k,a}\delta_{u+v,1}+(1-t_j)\delta_{k,a}\delta_{u+v,2} + t_j \delta_{u+v,2} \right) \Tr\left[\rho_{0} \left(G^{a}_{k}\right)_{u,v} \right] - \sum_{a,k}\Tr\left[G^{a}_{k} \Gamma^{a}_{k}\right] \\
+ & \sum_{i,k} \nu_{ki}\left(\Tr\left[ \rho_i \red{(G_k^a)_{0,0}} \right] - p(k|\rho_i)\right) + \Tr\left[P\left(\sum_{k}\red{(G_k^a)_{0,0}} - \mathbb{I}\right)\right] \\
+& \sum_{a,k}\sum_{u,v}\sum_{s=u+v}\Tr\left[Q_{s}^{a}\left(\left(G^{a}_{k}\right)_{u,v} - \frac{1}{D}\Tr\left[ \left(G^{a}_{k}\right)_{u,v} \right] \mathbb{I}\right)\right] ,
\end{aligned}
\end{equation}
where we introduced the dual variables $\Gamma^{a}_{k}$, $\nu_{ki}$, $P$ and $Q_{s}^{a}$ as Lagrangian multipliers. In the last term we also took into account that the last constraint from the primal holds for all elements $\left(G^{a}_{k}\right)_{u,v}$ that satisfy $u+v=s$, $\forall s$ (i.e., all blocks with the same sum of indices $u$ and $v$ are equivalent, for $s$ denotes the power of the scalar $z_{a}$).

The solution to the original SDP is always a saddle point of the Lagrangian $\mathcal{L}$, identified among the stationary points. The infimum of the Lagrangian over the variables of the primal semidefinite program reads
\begin{align}
\mathcal{I} = \underset{G^{a}_k}{\inf} \ \mathcal{L}  = c_m + \sum_{j=1}^{m-1} \underset{G^{a}_k}{\inf} \ \mathcal{L}_j  = c_m + \sum_{j=1}^{m-1} \underset{G^{a}_k}{\inf} \ \left\{\left(-\sum_{k,i}\nu_{ki}p(k|\rho_i) - \Tr\left[P\right]\right) + \sum_{a,k} \Tr\left[G_k^{a}K_k^{a}\right]\right\} \ ,
\end{align}
for
\begin{equation}
\begin{aligned}
	\left(K_k^{a}\right)_{0,0} &= \sum_{i}\nu_{ki}\rho_i + P - \sum_{a} \left(\Gamma_{k}^{a}\right)_{0,0} \\
	\left(K_k^{a}\right)_{u,v} &= 2\tau_j\rho_{0}\delta_{a,k} + Q_{1}^{a}\! -\! \frac{1}{D}\Tr\left[Q_{1}^{a}\right]\mathbb{I} - \sum_{u,v} \left(\Gamma_{k}^{a}\right)_{u,v} \quad \text{if} \ u+v=1 \\
	\left(K_k^{a}\right)_{u,v} &= \tau_j\rho_{0}\!\left[\left(1\!-\!t_j\right)\delta_{a,k}\!+\!t_j\right] + Q_{2}^{a}\! -\! \frac{1}{D}\Tr\left[Q_{2}^{a}\right]\mathbb{I} - \sum_{u,v} \left(\Gamma_{k}^{a}\right)_{u,v} \quad \text{if} \ u+v=2 \\
	\left(K_k^{a}\right)_{u,v} &= Q_{s}^{a}\! -\! \frac{1}{D}\Tr\left[Q_{s}^{a}\right]\mathbb{I} - \sum_{u,v} \left(\Gamma_{k}^{a}\right)_{u,v} \quad \text{if} \ u+v=s \ \text{for} \ s > 2
\end{aligned}
\end{equation}
collecting all terms in $\mathcal{L}_i$ multiplying the primal variables. The infimum $\mathcal{I}$ represents a lower bound on the optimal solution of the primal SDP. However, note that the infimum will diverge unless $K_k^{a}=0$, $\forall k,a$. To get good bounds therefore, we must maximise $\mathcal{I}$ over the set of dual variables that satisfy $K_k^{a}=0$. This maximisation leads to the formulation of the dual SDP,
\begin{equation}\label{eq:shannon_dual}
\begin{aligned}
   H(B|E) \geq c_m + \sum_{j=1}^{m-1} \ \underset{\Gamma^{a}_{k}, \nu_{ki}, P, Q_{s}^{a}}{\text{max}} & \quad \left(- \sum_{k,i}\nu_{ki}p(k|\rho_i) - \Tr\left[P\right]\right)  \\
\text{subject to} & \quad \Gamma_{k}^{a} \succeq 0 \\
& \quad \sum_{a} \left(\Gamma_{k}^{a}\right)_{0,0} = \sum_{i}\nu_{ki}\rho_i + P  \\
& \quad \sum_{u,v} \left(\Gamma_{k}^{a}\right)_{u,v} = 2\tau_j\rho_{0}\delta_{a,k} + Q_{1}^{a}\! -\! \frac{1}{D}\Tr\left[Q_{1}^{a}\right]\mathbb{I} \quad \text{if} \ u+v=1 \\
& \quad \sum_{u,v} \left(\Gamma_{k}^{a}\right)_{u,v} = \tau_j\rho_{0}\!\left[\left(1\!-\!t_j\right)\delta_{a,k}\!+\!t_j\right] + Q_{2}^{a}\! -\! \frac{1}{D}\Tr\left[Q_{2}^{a}\right]\mathbb{I} \quad \text{if} \ u+v=2 \\
& \quad \sum_{u,v} \left(\Gamma_{k}^{a}\right)_{u,v} = Q_{s}^{a}\! -\! \frac{1}{D}\Tr\left[Q_{s}^{a}\right]\mathbb{I} \quad \text{if} \ u+v=s \ \text{for} \ s > 2 \ .
\end{aligned}
\end{equation}
Any set of $rD\times rD$ (for $2r$ being the order of the SDP relaxation) positive semidefinite matrices $\Gamma_{k}^{a}$, $D\times D$ matrices $P$ and $Q_{s}^{a}$ and scalars $\nu_{ki}$ that satisfy the dual constraints above, represent a valid lower-bound on the Shannon entropy.

To recover the exact form as it is presented in the main text, one just needs to swap the maximisation and negative signs from the objective function, applying the identity $\max_x \left(-f(x)\right) = -\min_x f(x)$.


\section{An SDP for lower-bounding the single-round min-entropy}
\label{app.dual}

Here, we derive an SDP computing the single-round guessing probability and hence the min-entropy. The dual SDP does not feature the observed data in its constraints, only in the objective function and the solution provides an upper bound on the guessing probability for any set of data. As a result, the set of optimal parameters obtained by solving the dual problem for one set of data may be used to upper-bound the guessing probability for any other set of data by straightforward evaluation of the objective function.

\subsection{Primal SDP}

The min-entropy relative to Eve can be quantified by how well she is able to predict the outcomes in generation rounds, i.e.\ by the guessing probability $p_g$ that she correctly predicts the output $k$ when the input is $\rho_0$. Namely, by the left-over hash lemma, the asymptotic number of extractable random bits per round is then given by the min-entropy $\hm = -\log_2 p_g$. We now show how $p_g$ may be upper bounded (and hence $\hm$ lower bounded) by an SDP.

The measurement device implements some unknown, noisy measurement. While the user observes only the average behaviour, we assume Eve has perfect knowledge of these measurements, which we associate with a $d$-outcome POVM $\hat{\Pi}_k^\lambda$. Here, $\lambda$ labels the measurement strategies, which occur with distribution $q_\lambda$. For a particular POVM, the probability that Eve guesses correctly is determined by the most likely outcome $p_g = \underset{k}{\max} \Tr[\rho_0 \hat{\Pi}_k^\lambda]$. An upper bound on $p_g$ is then found by optimising over POVMs $\hat{\Pi}_k^\lambda$ and probability distributions $q_\lambda$, given the average observed behaviour
\begin{maxi}|s|[0]
{\{q_\lambda, \hat{\Pi}_k^\lambda\}}
{q_\lambda\max_k\Tr\left[\state{\rho}_0\hat{\Pi}_k^\lambda \right] \label{eq.pgoptim}}
{}
{p_g\leq}
\addConstraint{\sum_\lambda q_\lambda \Tr[\rho_i \hat{\Pi}_k^\lambda]}{=p(k|\rho_i)}{\ \ \forall i,k.}
\end{maxi}
This optimisation is not yet an SDP as it is nonlinear in  $\hat{\Pi}_k^\lambda$ and $q_\lambda$ and contains the maximisation over $k$. The latter issue can be resolved by observing that, while the number of strategies available to Eve is a priori unlimited, following the logic of Ref.~\cite{bancal2014more}, all strategies for which the most likely $k$ in \eqref{eq.pgoptim} is the same can be grouped together. Thus we need only consider $d$ different POVMs, and we can let $\lambda$ label the optimal $k$, i.e.~$\underset{k}{\max} \Tr[\state{\rho}_0\hat{\Pi}_k^\lambda] = \Tr[\state{\rho}_0\hat{\Pi}_\lambda^\lambda]$. The problem can now be linearised by introducing new variables $\hat{M}_k^\lambda = q_{\lambda}\hat{\Pi}_k^{\lambda}$. We obtain
\begin{maxi}
	  {\{\hat{M}_k^\lambda \}}{\sum_{\lambda=1}^{d}  \Tr\left[\state{\rho}_0\hat{M}_\lambda^{\lambda}\right]}{\label{eq.SDPprimal}}{p_g=}
	  \addConstraint{\hat{M}_k^\lambda \geq 0 \quad \forall k,\lambda}
	  \addConstraint{\sum_{k=1}^d \hat{M}_k^\lambda = \frac{1}{D}\Tr\left[\sum_k \hat{M}_k^\lambda\right] \mathbb{I}\quad \forall \lambda}
	  \addConstraint{\sum_{\lambda =1}^{d}\Tr[\rho_i\hat{M}_k^\lambda] = p(k|\rho_i) \ \ \forall i,k}
\end{maxi}
which is an SDP. 
Here, $D$ is the dimension of the POVM elements, and the first two constraints ensure that the $\hat{M}_k^\lambda$ form a valid POVM and the $q_\lambda$ a valid probability distribution. Note while ideally we do not want to constrain the dimension, when implementing the SDP, $D$ must be finite. $D$ should thus be chosen sufficiently large to not affect the optimum. Also note that in practice, it is often more useful to work with the dual SDP, in which the observed data enters in the objective function and not in the constraints.

\subsection{Dual SDP}

Firstly, recall the set of primal constraints \eqref{eq.SDPprimal}:
\begin{align}
    &\hat{M}_k^\lambda \geq 0 \quad \forall k,\lambda \in\{1, \dotsc, d\} \label{refreshConstraint1}\\
    &\sum_{k=1}^d \hat{M}_k^\lambda = \frac{1}{D}\Tr\left[\sum_k \hat{M}_k^\lambda\right] \mathbb{I}\quad \forall \lambda \in\{1, \dotsc, d\},
    \label{refreshConstraint2}\\
    &\sum_{\lambda=1}^{d} \Tr\left[\state{\rho}_i \hat{M}_k^\lambda\right]= p(k|\state{\rho}_i)\ \ \forall  i\in\{1, \dotsc, n\}, \, k\in\{1, \dotsc, d\}. \label{refreshConstraint3}
\end{align}
We begin by introducing a dual variable for each of the constraints above. For those defined by \eqref{refreshConstraint1}, we introduce semi-definite matrices $\{G_k^\lambda\}$. Similarly, for those defined by \eqref{refreshConstraint2} we introduce semi-definite matrices $\{H_\lambda\}$. Finally, for the constraints defined by \eqref{refreshConstraint3}, we introduce scalar variables $\{\nu_k^i\}$. Using these variables we form a Lagrangian from the sum of the primal objective function and, for each constraint, the product of the dual variable and the difference between the left- and right-hand side of the constraint. We obtain
\begin{align}
    \mathcal{L}&=\sum_\lambda \Tr\left[\state{\rho}_0 \hat{M}_\lambda^\lambda\right]\\
    &+ \sum_{k,\lambda} \Tr\left[G_k^\lambda \hat{M}_k^\lambda\right]\\
    & + \sum_\lambda\Tr\left[H_\lambda\left(\sum_k \hat{M}_k^\lambda - \frac{1}{D}\Tr\left[\sum_k \hat{M}_k^\lambda\right]\mathbb{I}\right)\right]\\
    &+\sum_{k,i} \nu_k^i\left(\sum_\lambda \Tr\left[\state{\rho}_{i}\hat{M}_k^\lambda\right] - p(k|\state{\rho}_i) \right).
\end{align}
Note that the last two terms vanish when the primal variables fulfill the primal constraints. Hence, maximising $\mathcal{L}$ provides an upper bound on $p_g$, if the $\{G_k^\lambda\}$ are chosen to be positive.  Now, we group all terms in which the primal SDP variables $\{\hat{M}_k^\lambda \}$ appear, thus we obtain
\begin{equation}
    \mathcal{L}=\sum_{\lambda, k} \Tr [\hat{M}_k^\lambda K_k^\lambda]-\sum_{k,i} \nu_k^i p(k|\state{\rho}_i) ,
\end{equation}
where we define
\begin{equation}
    K_k^\lambda:=\state{\rho}_0\delta_{k,\lambda} + G_k^\lambda + H_\lambda -\frac{1}{D}\Tr\big[H_\lambda\big]\mathbb{I} +
    \sum_{i=1}^n \nu_k^i \state{\rho}_{i}.
\end{equation}
Let $\mathcal{S}$ be the (unconstrained) supremum of the Lagrangian over the primal variables
\begin{equation}
    \mathcal{S}=\sup_{\{\hat{M}_k^\lambda \}} \mathcal{L} .
\end{equation}
One sees that, unless the $\{K_k^\lambda\}$ vanish, the supremum will diverge. Hence, to get a nontrivial upper bound, we demand that $K_k^\lambda=0$. Since the $\{G_k^\lambda\}$ are positive but otherwise arbitrary, this is equivalent to the condition
\begin{equation}
    \state{\rho}_0\delta_{k,\lambda}  + H_\lambda -\frac{1}{D}\Tr\big[H_\lambda\big]\mathbb{I} + \sum_{i=1}^n \nu_k^i \state{\rho}_{i}\leq 0.
\end{equation}
We thus arrive at the dual SDP by minimising the supremum
\begin{mini}
	  {\{H_\lambda\},\{\nu_k^i\}}{-\sum_{k=1}^d\sum_{i=1}^n\nu_k^i p(k|\state{\rho}_i)}{}{\bar{p}_g=}\label{eq.dualSDP}
	  \addConstraint{H_\lambda = H_\lambda^\dagger} \quad \forall \lambda ,
	  \addConstraint{\state{\rho}_0\delta_{k,\lambda}  + H_\lambda -\frac{1}{D}\Tr\big[H_\lambda\big]\mathbb{I} + \sum_{i=1}^n \nu_k^i \state{\rho}_{i}\leq 0 \quad \forall \lambda,k,}{}
\end{mini}
where the dual optimum bounds the primal as $p_g \leq \bar{p}_g$.

\section{Noisy, discretized measurement of a coherent state in the $X$-quadrature}
\label{app.cohprobs}

Here, we derive the probabilities associated with measuring the $X$-quadrature of a coherent state with homodyne detection subject to Gaussian excess noise followed by binning. Our starting point is the $X$-quadrature wavefunction of the coherent state of (generally complex) amplitude $\alpha$, given by
\begin{equation}
    \psi_\alpha(x) = \frac{1}{\pi^{1/4}}\exp\left[-\frac{1}{2}\left(x-\sqrt{2}\mathfrak{R}(\alpha)\right)^2 + i x \sqrt{2}\mathfrak{I}(\alpha)\right]
\end{equation}
where $\mathfrak{R}(\alpha)$ and $\mathfrak{I}(\alpha)$ are the real and imaginary parts of $\alpha$, respectively. Recalling the form of the average POVM operators from \eqref{eq.binnedhomPOVM}, we may compute the probabilities as
\begin{align}
  p(k|\alpha)=\bra{\alpha}\hat{\Sigma}_{k}\ket{\alpha} &= \bra{\alpha} \int_{I_k}\int_{-\infty}^\infty \ket{y}\bra{y}\frac{\exp\left[-(y-x)^2/2\noisevar\right]}{\sqrt{2\pi} \noisestd}\dd{y}\dd{x}\ket{\alpha}\\
  &= \int_{I_k}\int_{-\infty}^\infty \frac{\exp\left[-(y-x)^2/2\noisevar\right]}{\sqrt{2\pi}\noisestd} |\psi_\alpha(y)|^2\dd{y}\dd{x}\\
  &= \int_{I_k}\frac{1}{\sqrt{2\pi}\totstd}\exp\left[-\frac{\left(x-\sqrt{2}\mathfrak{R}(\alpha)\right)^2}{2\totvar}\right]\dd{x}\\ 
  &= \frac{1}{2}\left[\erf\left(\frac{b_k-\sqrt{2}\mathfrak{R}(\alpha)}{\sqrt{2}\totstd}\right)-\erf\left(\frac{a_k-\sqrt{2}\mathfrak{R}(\alpha)}{\sqrt{2}\totstd}\right)\right]
  \label{eq.homprob}
\end{align}
where $a_k$ ($b_k$) is the lower (upper) bound of the interval $I_k$, and $\totvar = \vacvar + \noisevar$ is the total variance of the output distribution in the $X$-quadrature before binning.

\section{Comparison of fixed and equal-probability bins}
\label{app.varbins}

As mentioned in the main text, it is interesting to consider the case in which the ADC bins may be of variable width such that the $ p(k|\state{\rho}_0)$ are all equal. In the case of homodyne measurements on coherent states, the probability associated with an arbitrary bin is given by \eqref{eq.homprob}. By asserting that all bins $d=2^\Delta$ are equiprobable for the vacuum input, we find that the interval $I_k$ of the $X$-quadrature covered by bin $k$ must be given by
\begin{equation}
    I_k = \sqrt{2}\totstd\bigg[\erf^{-1}\left(\frac{2k}{d}-1\right), \erf^{-1}\left(\frac{2(k+1)}{d}-1\right)\bigg]
\end{equation} 
with $k\in \{0,\dotsc,d-1\}$.

\begin{center}
\begin{figure*}[t]
    \includegraphics[width=0.5\linewidth]{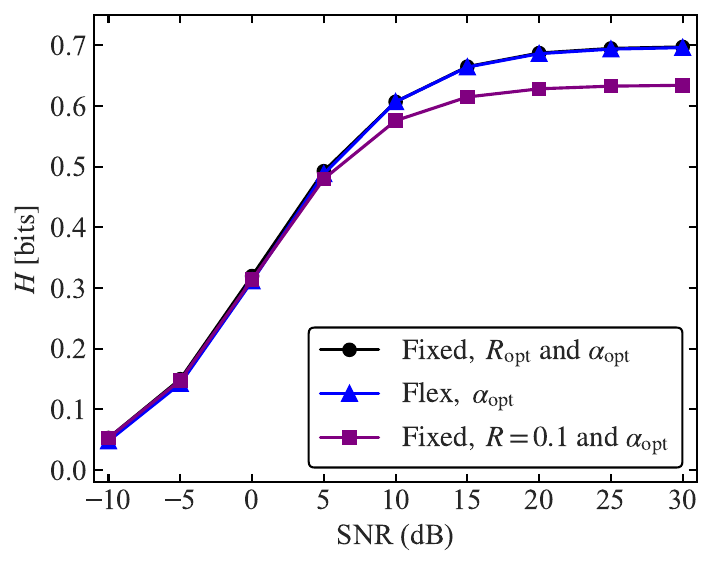}
	\caption{Comparison of fixed- vs flexible-bin ADC configurations for the coherent state protocol.}
	\label{fig.varvsfixedbin}
\end{figure*}
\end{center}

In \figref{fig.varvsfixedbin}, we compare the entropy of the coherent state protocol with fixed and flexible ADCs. As the coherent state approach requires multiple output distributions, changes to the entropy associated with a flexible- vs fixed-bin ADC are non-trivial. We consider three realisations of the protocol: 
\begin{enumerate}
    \item A fixed-bin ADC with $2^2$ bins. Selecting range $R=R_\text{opt}$ and probe state amplitudes $\{0, \alpha_\text{opt}\}$ where $R_\text{opt}$ and $\alpha_\text{opt}$ are the values of $R$ and $\alpha$ that maximise the rate at any given SNR.
    \item A flexible-bin ADC with $2^2$ bins and probe state amplitudes $\{0, \alpha_\text{opt}\}$
    \item A fixed-bin ADC with naive range $R=0.1$ and probe states $\{0, \alpha_\text{opt}\}$\label{item:scenario3}
\end{enumerate}
In each case we assume ideal parameters $\eta=1$ and $\gamma=0$. We find that, at any SNR, there are optimal choices of $R$ and $\alpha$ under which the fixed-bin ADC in scenario (1) outperforms the flexible-bin ADC in scenario (2). Despite this, we find that realisation (2) outperforms realisation (3) for the naive choice of $R=0.1$ as shown in \figref{fig.varvsfixedbin}. These results suggest that a fixed-bin ADC is optimal if the SNR is known to a good degree of certainty, and the ADC range and coherent state amplitudes can be finely tuned. If any of these conditions are not satisfied, a flexible-bin ADC, offering consistently high entropy under a broad range of parameters, would be preferred.

\section{Considerations regarding implementation of the SDP in practice}
\label{app.SDPdetails}

Here, we consider the finite-size effect in the observed distributions and the effect of the chosen Fock-space cutoff on the SDP results. We also briefly discuss the software used to obtain our results.

\subsection{Finite-size effects}
\label{sec:finite_size}
In our protocol, we evaluate the amount of randomness which can be generated from measurement results based on experimentally observed probability distributions. However, when the number of experimental rounds is finite, we will obtain frequencies instead of probability distributions. Let $N$ represent the number of total experimental rounds and $N_{k, i}$ denote the number of events where we obtain measurement outcome $k$ given that the probe state was $\rho_i$. Then, we obtain the following frequencies $\text{freq.} (k|\rho_i) = N_{k, i}/\sum_{k}N_{k, i}$.

To evaluate the finite-size effect, a relevant parameter is the $\epsilon$-smooth min-entropy accumulated during the protocol $H_\text{min}^\epsilon (B^N|E^N)$~\cite{Renner2004}. We utilize the generalized entropy accumulation theorem (GEAT)~\cite{metger2022, Metger2023} to bound the smooth min-entropy. Roughly speaking, in GEAT, we can lower bound the smooth min-entropy with the single-round von Neumann entropy up to some constant values. Indeed, for $\epsilon \in (0,1)$ and $\alpha \in (1, 3/2)$, the GEAT can be expressed as follows
\begin{equation}\label{eq:geat}
    \frac{1}{N}H^\epsilon_\text{min} (B^N|E^N) \geq \min_{C^N \in \Omega}  f_\text{min} (\text{freq.} (C^N)) - \frac{\alpha -1}{2 - \alpha} \frac{\ln (2)}{2} V^2 - \frac{1}{N} \frac{g(\epsilon) + \alpha \log (1/ p_\Omega)}{\alpha - 1} - (\frac{\alpha -1 }{2 - \alpha})^2 K'(\alpha).
\end{equation}
In Eq.~(\ref{eq:geat}), $f_\text{min}$ is a so-called min-tradeoff function which is a function of frequencies and gives a lower bound on the single-round conditional von Neumann entropy, i.e.,
\begin{equation}
    f_\text{min} (q) \leq \min_{\nu \in \sum_n (q)} H(B_n|E_n)_\nu,
\end{equation}
where $q$ represents observed frequencies and $\sum_n (q)$ is the set of states that can be produced after the measurement in the $n$th round corresponding to $q$. The minimisation of the min-tradeoff function in Eq.~(\ref{eq:geat}) is performed over all events which belong to some specific event $\Omega$ occurring with probability $p_\Omega$. The other parameters in \eqref{eq:geat} can be expressed as follows:
\begin{align}
    g(\epsilon) &= - \log(1- \sqrt{1 - \epsilon^2}),\\
    V &= \log (2 d^2 + 1) + \sqrt{2 + \text{Var} (f_\text{min})},\\
    K'(\alpha) &= \frac{(2-\alpha)^3}{6(3-2\alpha)^3 \ln (2)} 2^{\frac{\alpha-1}{2-\alpha} (2 \log d + \text{Max} (f_\text{min}) - \text{Min} (f_\text{min}))} \ln^3 (2^{2 \log d + \text{Max} (f_\text{min}) - \text{Min} (f_\text{min})} + e^2),
\end{align}
where $d$ is the number of measurement outcomes and $\text{Max} (f_\text{min})$, $\text{Min} (f_\text{min})$ and $\text{Var} (f_\text{min})$ represent the maximum, the minimum and the variance of the min-tradeoff function $f_\text{min}$, respectively.

To apply the GEAT, it is important to choose an appropriate min-tradeoff function. In this work, we use the objective function of the dual SDP in Eq.~(\ref{eq:shannon_dual}) as a min-tradeoff function:
\begin{equation}
    f_\text{min} (\text{freq.} (k|\rho_i)) = c_m - \sum_{j = 1}^{m-1} \left( \sum_{k, i} \nu_{ki}^* \text{freq.} (k|\rho_i) + \text{Tr} [P^*]  \right),
\end{equation}
where $\nu_{ki}^*$ and $P^*$ represent the optimal parameters at each iteration $j$. From this min-tradeoff function, we need to calculate its maximum, minimum and variance. To this end, we take the following procedure. First, we solve the dual SDP for frequencies calculated from the whole experimental data, and keep values of $\nu_{ki}^*$ and $P^*$. Then, we divide the whole experimental data into several subsets. For each subset, we calculate frequencies and values of the min-tradeoff function using the values of $\nu_{ki}^*$ and $P^*$. From the prepared values of the min-tradeoff function, we calculate the maximum, minimum and variance. Furthermore, in our calculation, $\Omega$ corresponds to events where we certify randomness and we set $\alpha = 1 + 1/\sqrt{N}$.

\subsection{Fock-space cutoff}

In order to implement the SDP \eqref{eq:shannon_dual} in practice, the variables must be represented by finite-dimensional matrices. For an optical protocol, it is natural to work in Fock space. However, the coherent states used as probe states in the specific protocol considered here have support on the entire infinite-dimensional space. Hence, a cutoff must be applied.

For the POVM  elements in the SDP which we solve to evaluate the amount of randomness $\hat{\Pi}_k$ and a probe state $\rho_i = \ketbra{\alpha_i}{\alpha_i}$, the following holds
\begin{equation}
  \bra{\alpha_i} \hat{\Pi}_k\ket{\alpha_i} = p(k|\rho_i) ,   
\end{equation}
where $p(k|\rho_i)$ denotes observed probability distribution. Expressing this in the Fock basis we obtain
\begin{equation}
  e^{-|\alpha_i|^2}\sum_{\mu=0}^\infty\sum_{\nu=0}^\infty\frac{(\alpha_i^*)^\mu\alpha_i^\nu}{\sqrt{\mu!\nu!}}\bra{\mu}\hat{\Pi}_k\ket{\nu} = p(k|\alpha_i) .
\end{equation}
When implementing the SDP in practice, we truncate the infinite sums appearing here by introducing a cutoff $D$ in Fock space
\begin{equation}
   e^{-|\alpha_i|^2}\sum_{\mu=0}^{D-1}\sum_{\nu=0}^{D-1}\frac{(\alpha_i^*)^\mu\alpha_i^\nu}{\sqrt{\mu!\nu!}}\bra{\mu}\hat{\Pi}_k\ket{\nu} = p(k|\alpha_i) .
   \label{eq.pobsFinite}
\end{equation}
The $\{\hat{\Pi}_k\}$ are taken to live in this $D$ dimensional space.

\begin{figure}
    \includegraphics[width=0.5\linewidth]{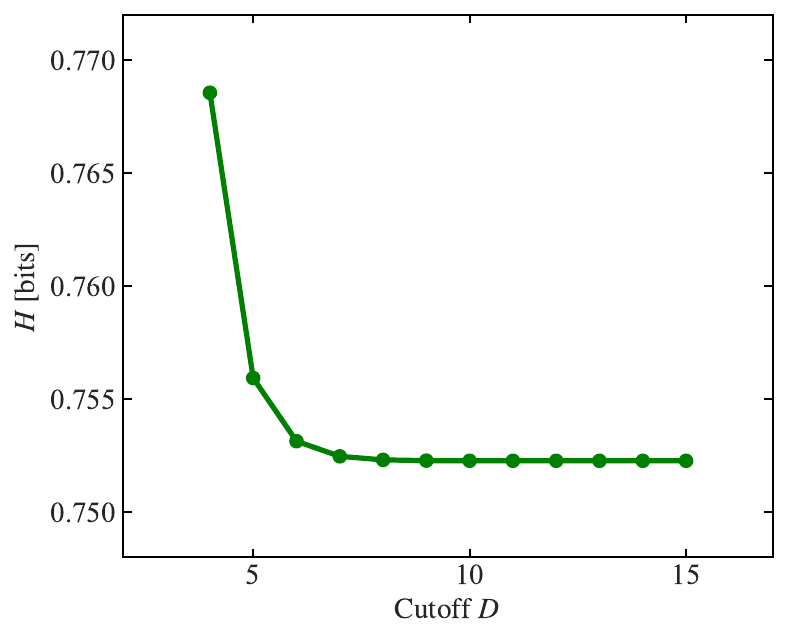}
	\caption{Convergence of the von Neumann entropy with Fock-space cutoff $D$. We use 4 probe states with amplitudes $[-0.002, -0.447, 0.902, -1.324]$, $\Delta=2, R = 0.5, \gamma = 0$, and $\eta = 1$.}
	\label{fig.fockconvergence}
\end{figure}

Here, the cutoff must be sufficiently large to allow the construction of operators on the left-hand side that are physically able to reproduce observed distributions, otherwise the SDP becomes infeasible. Furthermore, too small cutoff may limit Eve's strategies and lead to overestimation of randomness. In \figref{fig.fockconvergence}, we plot the conditional von Neumann entropy against the Fock-space cutoff $D$. We observe that the entropy decreases as the cutoff increases and it saturates when we set $D$ larger than 8.

\subsection{Software}

For the results throughout this work, all SDPs were implemented in Python using the \texttt{CVXPY} package~\cite{diamond2016cvxpy, agrawal2018rewriting} to express the problem programmatically in a high-level form before obtaining the solution using the \texttt{MOSEK} conic optimization package~\cite{Mosek2024}.

\section{Experimental setup}
\label{sec:app_exp}

\begin{figure}[t]
	\centering
	\includegraphics[width=\textwidth]{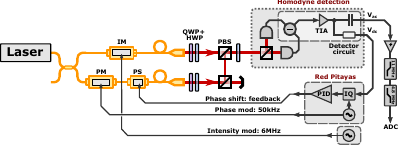}	
	\caption{\label{fig:exp_setup}Experimental setup to generate and measure a sideband coherent state. The components used are 1550 nm fiber laser (NKT Photonics Koheras BASIK E15), fiber intensity modulator (IM, iXblue MPZ-LN-10) , fiber phase modulator (PM, iXblue MPX-LN-0.1), fiber phase shifter (PS, General Photonics FPS-002), quarter-wave plate (QWP), half-wave plate (HWP), polarization beamsplitter (PBS), polarization based homodyne detector with a trans-impedance amplifier (TIA) stage and AC/DC voltage splitting and an analogue-to-digital converter (ADC, GaGe Razor Express16XX CompuScope PCIe digitizer board).}
\end{figure}

Here we present the experimental setup, illustrated in \figref{fig:exp_setup}. A coherent state is generated at a sideband frequency by intensity modulating a seed beam. The amplitude of the coherent state can be controlled by adjusting the peak-to-peak voltage of the signal driving the modulator. The state is measured by homodyne detection, that is locked to the $X$-quadrature of the seed beam. The lock is implemented using a standard AC locking scheme; the optical local oscillator is phase modulated at \SI{50}{\kHz} by an EOM and the DC signal of the homodyne detector is then downmixed at \SI{50}{\kHz} and lowpass filtered to reveal the error signal. The error signal is then passed to a PID controller that locks the phase of the LO to be in phase with the seed beam and therefore also the coherent sideband state. The AC output of the homodyne detector, carrying the quadrature information of the coherent sideband state is then amplified and bandpass filtered before being digitized by a 16-bit ADC.

\subsection{Digital signal processing chain}
\label{sec:app_dsp}
Here we describe in detail the four parts of the first block of the digital signal processing (DSP) pipeline. The first part is the digital downmixing of the raw 16-bit homodyne time traces $V(t)$, which in itself consist of three steps: 1) finding the downmixing frequency, 2) finding the downmixing phase and 3) downmixing. The second part is conversion from bit values to quadrature values and the final two parts are estimation of the corresponding amplitudes of the measured coherent probe states and construction of downsampled quadrature distributions to be passed to the SDP protocol.

\begin{figure}[t]
	\centering
\includegraphics[width=\textwidth]{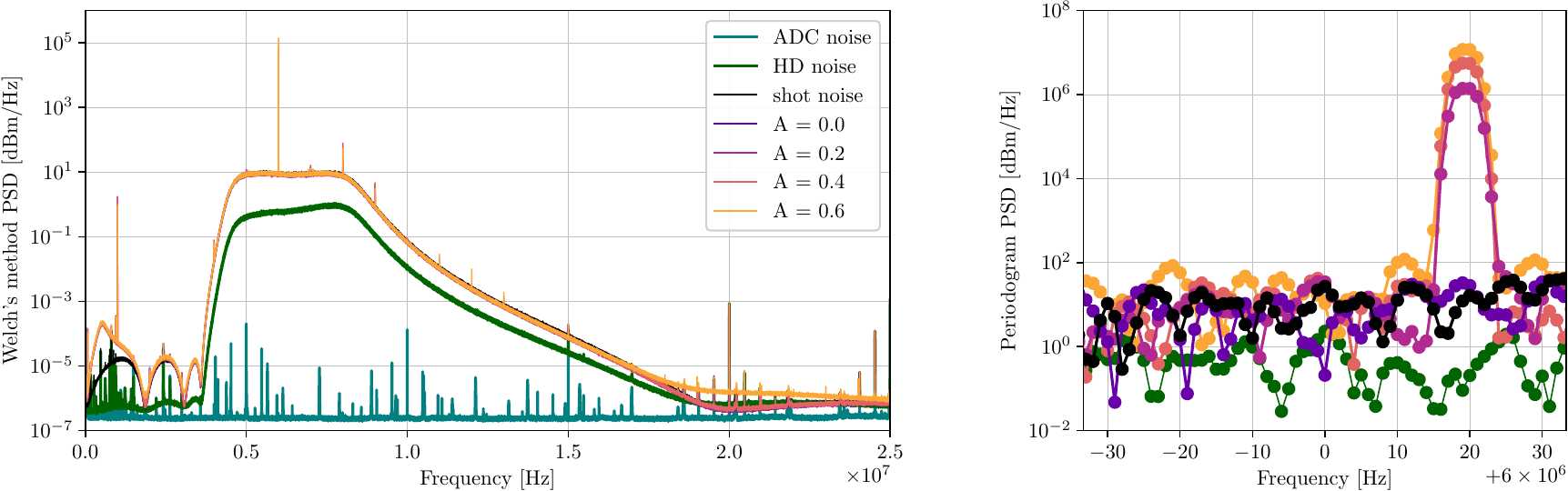}
	\caption{\label{fig:exp_dsp} \textbf{(left)} PSDs of the raw 16-bit homodyne time traces calculated using Welch's method, including electronic noise traces from the ADC and homodyne detector (HD). Welch's method results in filtered traces that better visualize the structure of the full frequency spectrum. The sideband modulation peak is clearly seen at \SI{6}{\MHz}, together with the bandpass filtering effect of the \SI{11}{\MHz} LPF and \SI{4.8}{\MHz} HPF. \textbf{(right)} Zoom of the PSDs  around \SI{6}{\MHz} calculated using periodograms to avoid any filtering of the peak structure. The peak height is seen to correlate correctly with the modulation strength and that there is a sufficient clearance to the HD noise floor.}
\end{figure}

\figref{fig:exp_dsp} shows plots of the power spectral density (PSD) of the raw time traces. Note that here and below, the data set used for plotting is different from the one in the main text (which contains more datapoints), but the methods and procedures are exactly the same.

\subsection{Digital downmixing}
\label{sec:app_dsp1}
Downmixing is typically performed in the electrical domain by mixing the signal with a corresponding electrical modulation signal followed by low-pass filtering (LPF) in order to convert the sideband signal to a DC value. In the digital domain the process is similar and the downmixed time traces are calculated as follows:
\begin{equation}
v_\text{dm}(t) = f_\text{lpf} \left[v(t) \mathfrak{R}(v_\text{dlo}(t))\right] + if_\text{lpf} \left[v(t) \mathfrak{I}(v_\text{dlo}(t))\right]
\end{equation}
where $v(t)$ is the input signal, $v_\text{dlo}(t) = e^{i(2\pi t f_\text{dlo}+\phi_\text{dlo})}$ is the digital local oscillator (DLO) and $f_\text{lpf}$ is a filter function that implements a linear digital filter twice, once forward and once backwards, using a 501st-order low pass finite impulse response (FIR) filter with a cutoff frequency of $f_c=\SI{1}{\MHz}$. Both $f_\text{lpf}$ and the FIR filter are implemented with the \texttt{scipy.signal} python package using the \texttt{filtfilt} and \texttt{firwin} functions, respectively~\cite{2020SciPy-NMeth}. But before the downmixing can be performed we first need to determine the proper frequency $f_\text{dlo}$ and phase $\phi_\text{dlo}$ of the DLO.\\

\begin{figure}[t]
	\centering
	\includegraphics[width=\textwidth]{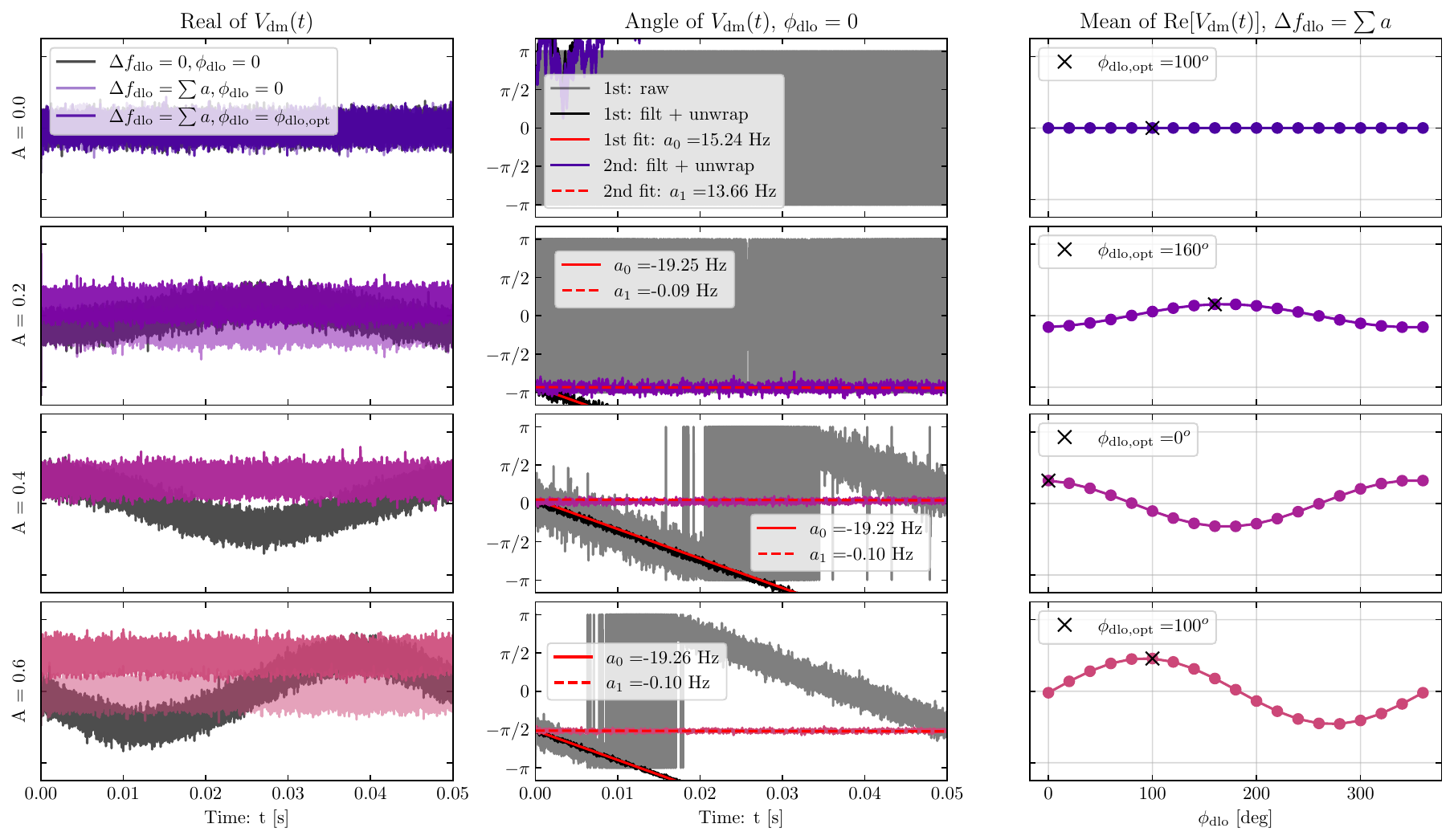}
	\caption{Overview of the steps involved in the digital downmixing procedure.}
	\label{fig:app_dsp1}
\end{figure}

\subsubsection{Finding the DLO frequency offset: $\Delta f_\text{dlo}$}
Even though we set the function generator to provide a $f_\text{mod}=\SI{6}{\MHz}$ signal to the amplitude modulator the actual frequency of the optical signal will vary slightly. We therefore need to find the frequency offset $\Delta f_\text{dlo}$ needed to adjust the DLO to match the actual frequency of the optical signal. This offset can be estimated by fitting the evolution of the angle of the complex valued $v_\text{dm}(t)$, as the frequency mismatch between the DLO and signal will result in a constant phase drift of the downmixed signal. A visualization of this is shown in \figref{fig:app_dsp1}. Here the left column shows the drift of the downmixed signal (in grey) when $f_\text{dlo}= f_\text{mod}$, the center column shows the corresponding angle (in grey) of $V_\text{dm}(t)$, which after filtering and unwrapping can be fitted with $a t+b$. A new downmixed signal is then calculated with $f_\text{dlo} = f_\text{mod} + \Delta f_\text{dlo}$, where $\Delta f_\text{dlo}=a_0$ and from that trace a second smaller correction term $a_1$ is estimated. The process is repeated $n$ times with $\Delta f_\text{dlo} = \sum_n a_n$ until the resulting phase drift is fully eliminated ($a_n\approx 0$). The result of this intermediate step is represented by the transparent traces in the left column of \figref{fig:app_dsp1}. We note that for our data, only a single step was required as $a_1\approx 0$ for all probe states and that the same DLO offset could be obtained independent of the chosen DLO phase.

\subsubsection{Finding the optimal downmixing phase: $\phi_\text{dlo,opt}$} 
Using the correct DLO frequency will result in a downmixed time trace of constant mean value. This value then depends on the phase of the DLO, $\phi_\text{dlo}$, relative to the signal. We therefore need to find the $\phi_\text{dlo,opt}$ that maximises the obtained mean value of the downmixed time trace. This can be done by either a simple brute force search or optimization routine. Such a search is shown in the right column of \figref{fig:app_dsp1} and the resulting proper downmixed signals as the slightly transparent traces in the left column. It is worth noting that shifting $\phi_\text{dlo,opt}$ by $180^o$ then corresponds to having measured the same probe state with the phase between the optical LO and probe state rotated by $180^o$.

\subsection{Bit-to-quadrature value conversion}
\label{sec:app_dsp2}
In order to convert the unit of time traces from bits (or voltage) into quadrature values, the trace is normalized against the shot noise level (SNL). A normalized time trace is therefore calculated as $\hat{x} = V(t) / (\sqrt{2} V_\text{snl})$, where $V_\text{snl}$ is the SNL and is obtained as the standard deviation of the shot noise trace. Using this normalization the variance of the shot noise trace becomes equal to the vacuum variance $\langle \Delta \hat{x}^2_\text{sn}\rangle = 1/2$. The normalization of the downmixed traces is then naturally performed using the SNL of the corresponding downmixed shot-noise trace and the result is shown in the upper row of \figref{fig:app_dsp3}.

\subsection{Estimation of probe state amplitudes}
\label{sec:app_dsp3}
\begin{figure}[t]
	\centering
    \includegraphics[width=0.9\textwidth]{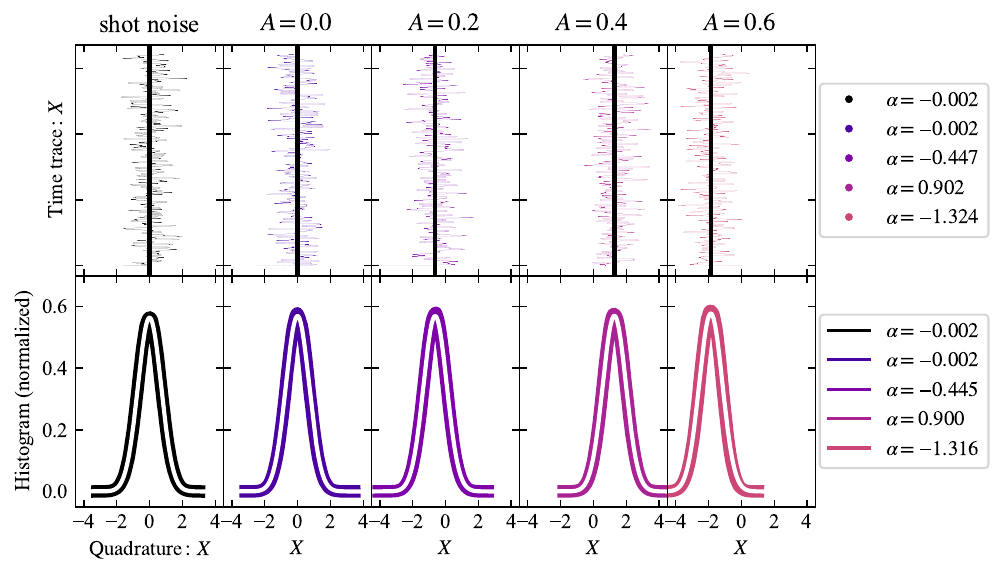}
	\caption{\textbf{(top row)} Normalized downmixed time traces. The vertical black lines indicate the position of the mean and its value is given by legend. \textbf{(bottom row)} $X$-quadrature distributions of the measured probe states calculated as normalized 500 bin histograms in the quadrature range [-4.5, 4.5]. The white/black lines indicate the best fit of $\mathrm{Pr}_{\alpha}(x)$ to the histogram data and the corresponding amplitude is given by the legend.}
	\label{fig:app_dsp3}
\end{figure}

Given the normalization and that we employ coherent states of real amplitude only, the amplitude $\alpha$ of any such state is equal to the average quadrature value corresponding to taking the mean of the normalized downmixed time trace: $\alpha = \langle\hat{x}\rangle$. Alternatively, using a model of the expectation value of a coherent state $\mathrm{Pr}_{\alpha}(x)= \pi^{-1/2}  e^{-(x - \sqrt{2}\alpha)^2}$, the value of alpha that best fits the measured quadrature distribution (histogram of $\hat{x}$ values) is found. The results of both methods are shown in \figref{fig:app_dsp3} and show good agreement.

\subsection{Downsampling}
\label{sec:app_dsp4}
The ADC we have used in our setup has a depth of 16 bits. However, solving the SDP with $2^{16}$ constraints for each probe state would require a prohibitive runtime even with generous computational resources. Moreover, as shown in the main text, increasing the bit depth offers little improvement to the extractable randomness after the bit depth exceeds $\Delta=4$ under realistic parameters. Therefore, we perform downsampling in order to create new histograms with a smaller bit depth over an adjustable range less than the experimental range of the ADC and thus we can solve the SDP rapidly with little reduction in the output entropy.

\begin{figure}[h!]
	\centering
    \includegraphics[width=\textwidth]{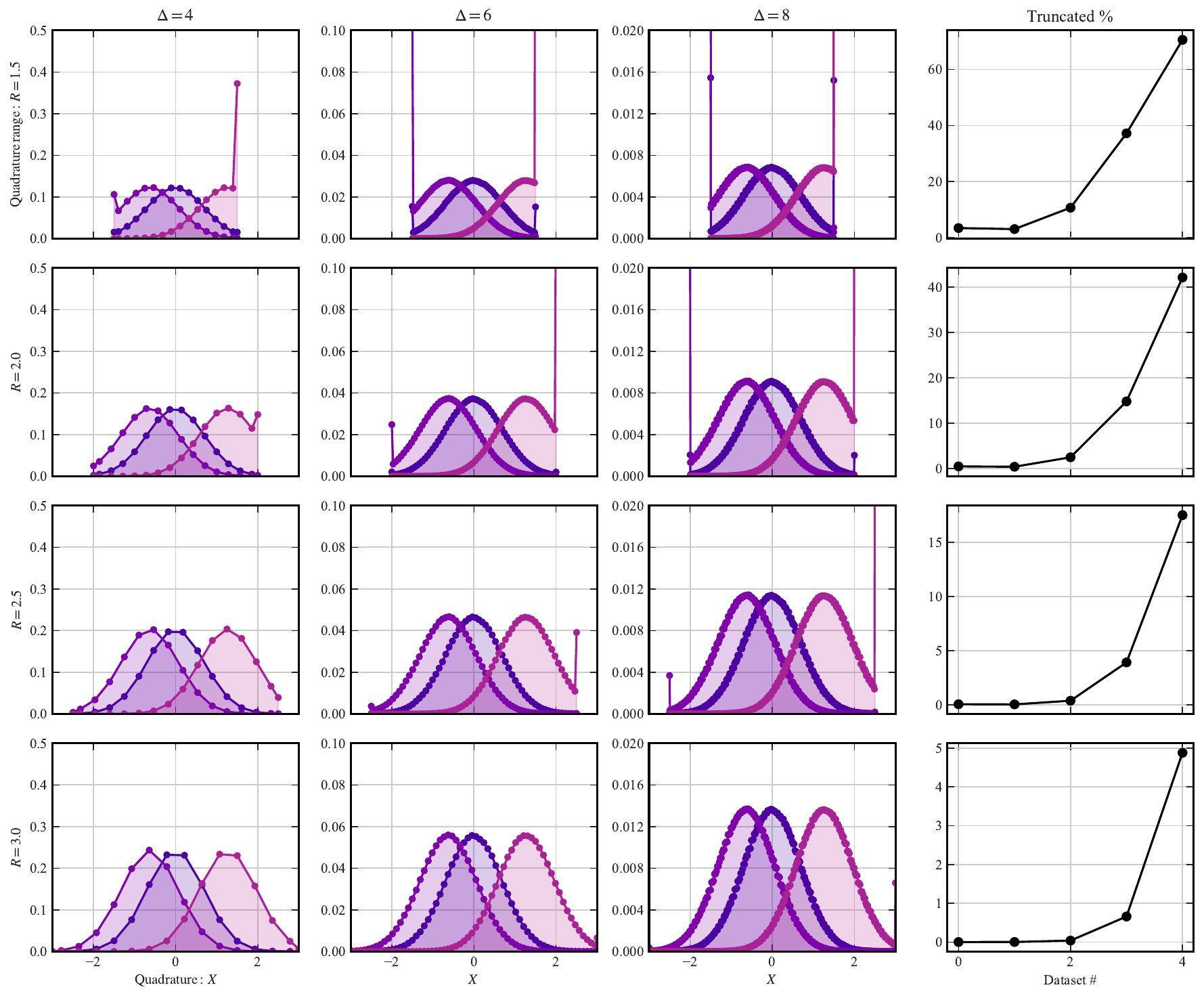}
	\caption{Histograms of downsampled versions of the normalized downmixed time traces from the top row of \figref{fig:app_dsp3}. The y-axis of the histograms corresponds to bin frequencies. The percentage of data points truncated by the corresponding downsampling is shown in the rightmost plots, with data set number corresponding to the sequence [shot noise, $A=0.0$, $A=0.2$, $A=0.4$, $A=0.6$].}
	\label{fig:app_dsp4}
\end{figure}

To illustrate the downsampling procedure, we consider bit depths of $\Delta=4$, $6$ and $8$ and quadrature ranges of $R=1.5$, $2.0$, $2.5$ and $3.0$. The normalized downmixed traces are binned under the definition of the ADC bin function given by \eqref{eq.fixedBinFunction} in the main text for a given $R$ and $\Delta$. The downsampled histograms are shown in \figref{fig:app_dsp4}. Bin frequencies are presented as points plotted at the center of all interior bins while points at $-R$ and $R$ show the frequencies of the end bins at the lower- and upper-end of the range, respectively. The percentage of the total data points that fall into either of the end bins is shown for each dataset on the right-hand side for each $R$ and is independent of the chosen bit depth.

\section{Effect of amplitude estimation errors}
\label{app.amplitudeerr}

In any experimental implementation, there is necessarily some uncertainty in the amplitudes of the trusted probe states. As the probe states are assumed to be known in the entropy certification step, such uncertainties may, in principle, compromise the protocol's security, i.e.\ lead to overestimation of the entropy.
    
Intuitively, if actual probe states are more distinguishable than assumed, Eve can cheat. Indeed, if the probe states are perfectly distinguishable, Eve can cheat by first identifying the state and then outputting according to any local probability distribution she likes. The constraints (in the SDP) that the observed distributions must be reproduced do not then constrain her measurement in the generation rounds in any way, and there is no randomness relative to Eve.

\begin{figure}[t]
    \includegraphics[width=0.65\linewidth]{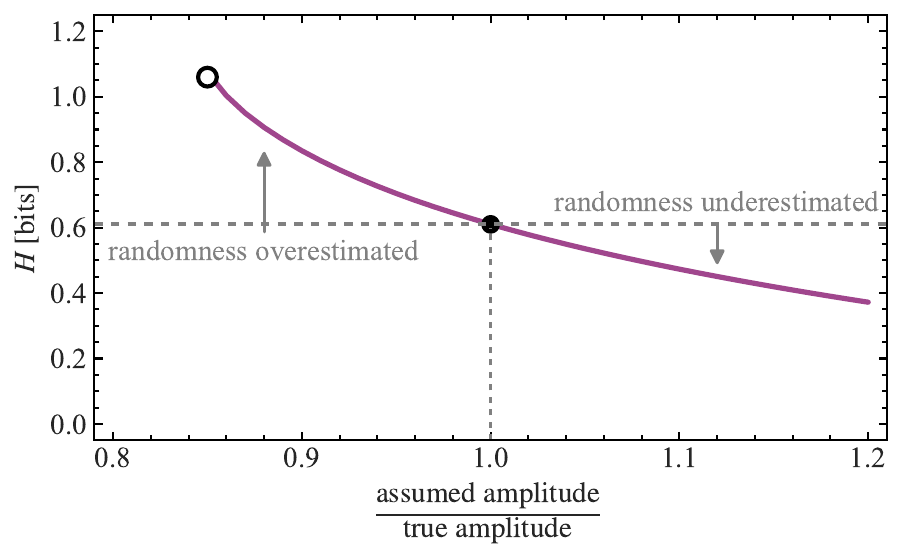}
	\caption{Von Neumann entropy (as computed via \eqref{eq:shannon_obj}) vs. amplitude estimation error, for the coherent-state protocol using four states with amplitudes $\alpha_0 = 0$, $\alpha_1 = 0.33$, $\alpha_2 = 0.66$, $\alpha_3 = 1.0$, and $\Delta=2$, $R=1.5$, $\gamma=0$, $\eta=1$, and an SNR of \SI{15}{\dB}. The filled circle and horizontal, dashed line indicate the true value of the von Neumann entropy corresponding to no estimation error. To the left of the open circle, the SDP becomes infeasible (it becomes impossible to reproduce the true distributions by measurements on the less distinguishable assumed coherent states).}
	\label{fig.wrongalphaHmin}
\end{figure}

In \figref{fig.wrongalphaHmin}, we investigate the effect of the discrepancy between the assumed and true probe-state amplitudes. We fix the values of the true amplitudes that enter in the distributions $p^\gamma(k|\alpha)$, and vary the assumed amplitudes. We take all the amplitudes to be off by the same factor, i.e.\ we take $\alpha_i^\text{assumed} = r \alpha_i^\text{true}$ for all $i$ and vary $r$. The plot shows the von Neumann entropy as computed via the SDP vs.\ the scaling factor.

We see that, as expected based on the intuition above, overestimation of the amplitudes leads to underestimation of the entropy. In this regime, the protocol is thus still secure, but sub-optimal. Underestimating the amplitudes, however, leads to an overestimation of the entropy, compromising security. Hence, to ensure that amplitude estimation uncertainty does not lead to security holes, one should pick assumed values for the SDP which err to the side of larger values with respect to the estimation.

\end{document}